\documentclass[twoside,11pt]{article}
\usepackage{blindtext}
\usepackage{jmlr2e}

\usepackage{amsmath}  
\usepackage{graphicx}
\usepackage{booktabs}
\usepackage{makecell}
\usepackage{amsfonts}
\usepackage{subcaption}
\usepackage{listings}
\usepackage{tikz}
\usepackage{xcolor}
\usepackage{enumitem}
\usepackage{colortbl}
\usepackage{bbm}
\usepackage{pdflscape}

\usetikzlibrary{positioning, arrows.meta}
\usetikzlibrary{calc}

\DeclareMathOperator*{\argmax}{arg\,max}

\usepackage{bbm}

\newtheorem{assumption}{Assumption}

\usepackage{lastpage}

\ShortHeadings{Automatic Differentiation of Agent-Based Models}{Quera-Bofarull et al.}
\firstpageno{1}

\begin{document}

\title{Automatic Differentiation of Agent-Based Models}

\author{\name Arnau Quera-Bofarull \email arnauq@proton.me \\
       \addr Macrocosm and University of Oxford\\
       \name Nicholas Bishop \email nicholas.bishop@cs.ox.ac.uk\\
       \name Joel Dyer \email joel.dyer@cs.ox.ac.uk \\ 
       \name Daniel Jarne Ornia \email daniel.jarneornia@cs.ox.ac.uk \\
       \name Anisoara Calinescu \email ani.calinescu@cs.ox.ac.uk \\
       \name Doyne Farmer \email doyne.farmer@smithschool.ox.ac.uk \\
       \name Michael Wooldridge \email mike.wooldridge@cs.ox.ac.uk \\
       \addr University of Oxford} 

\editor{My editor}

\maketitle

\begin{abstract}

    Agent-based models (ABMs) simulate complex systems by capturing the bottom-up interactions of individual agents comprising the system. Many complex systems of interest, such as epidemics or financial markets, involve thousands or even millions of agents. Consequently, ABMs often become computationally demanding and rely on the calibration of numerous free parameters, which has significantly hindered their widespread adoption. In this paper, we demonstrate that automatic differentiation (AD) techniques can effectively alleviate these computational burdens. By applying AD to ABMs, the gradients of the simulator become readily available, greatly facilitating essential tasks such as calibration and sensitivity analysis. Specifically, we show how AD enables variational inference (VI) techniques for efficient parameter calibration. Our experiments demonstrate substantial performance improvements and computational savings using VI on three prominent ABMs: Axtell’s model of firms; Sugarscape; and the SIR epidemiological model. Our approach thus significantly enhances the practicality and scalability of ABMs for studying complex systems.
    
\end{abstract}


\section{Introduction}
\label{sec:intro}
An agent-based model (ABM) aims to simulate a complex system in a \emph{bottom-up} fashion, by explicitly modelling the behaviour of autonomous agents who collectively comprise the system. The power of ABMs lies in their ability to capture and explain how macroscopic behaviour emerges naturally from the interactions of individual components. This contrasts sharply with \emph{top-down} methods, which model macroscopic phenomena directly without considering underlying mechanisms \citep{van_dyke_parunak_agent-based_1998, rahmandad_heterogeneity_2008}.

For example, consider modelling the spread of an epidemic throughout a city. An ABM explicitly models the infection status of each citizen and their social interactions. Aggregating individual infection statuses through time recovers an infection curve describing how total infections vary over time. In contrast, a top-down approach may model the infection curve directly through differential equations. While both methodologies can reproduce historical infection curves, only the ABM reveals the underlying mechanisms by which they arise from individual social behaviour. This ability to model emergent macroscopic phenomena has made agent-based modelling a valuable tool across diverse fields, including economics \citep{axtell_agent-based_2025}, epidemiology \citep{tracy_agent-based_2018}, ecology \citep{grimm_individual-based_2005}, and the social sciences \citep{macy_factors_2002}.

Despite their expressive power, ABMs raise significant computational challenges that limit their practical application. In this paper, we focus on two key issues: \emph{calibration} and \emph{sensitivity analysis}. Calibration involves fine-tuning parameter values so that ABM outputs match real-world empirical observations \citep{platt_comparison_2020, mcculloch_calibrating_2022, dyer_black-box_2024}. Due to the expressiveness of typical ABMs, multiple parameter configurations often reproduce the available empirical data. For instance, different demographics or social behaviours could explain disease spread during an epidemic, while various market conditions might trigger an economic crash \citep{paulin_understanding_2019}. Beyond reproducing complex system behaviour, researchers often need to assess how sensitive a system is to small changes in conditions. For example, stock exchanges like Nasdaq want to understand how changes to tick size influence trading behaviour, while climate policymakers seek sensitive intervention points where small policy changes can drive transformative shifts in fossil fuel usage. Agent-based modelling provides a natural framework for studying such sensitivity by maintaining high-fidelity representations at the individual agent level. Understanding which parameters an ABM is most sensitive to thus provides insights into the sensitivity of the corresponding real-world system. However, the high-dimensional parameter spaces and computational cost of ABMs make sensitivity analysis a significant challenge \citep{edeling_impact_2021}.

These computational challenges have motivated us to consider the following question: \emph{Can automatic differentiation be applied to ABMs to yield similar computational benefits as it has for deep learning?} While the uptake of agent-based modelling has been relatively slow, the adoption of machine learning methods, especially those based in \emph{deep learning}, has exploded across both industry and academia. Automatic differentiation (AD) \citep{baydin_automatic_2018} has been crucial to addressing deep learning's computational challenges by enabling efficient gradient computation through complex computational graphs, allowing neural architectures to be trained via first-order methods. The potential impact for ABMs is substantial: if gradients of ABM outputs with respect to parameters were available, local sensitivity analysis could be performed efficiently in a single simulation run, rather than requiring many expensive parameter perturbations \citep{quera-bofarull_dont_2023}. Moreover, gradient information could enable more efficient calibration procedures through gradient-based optimization techniques \citep{chopra_differentiable_2023, dyer_gradient-assisted_2023}. However, applying AD to ABMs is non-trivial due to their highly discrete and stochastic nature—agents make discrete choices from finite action sets, and environments evolve randomly based on their decisions.

For calibration specifically, procedures that provide \emph{uncertainty quantification} over parameter values are desirable, as they provide the modeller with a complete picture regarding the potential causes of emergent macroscopic behaviours. With this in mind, Bayesian inference procedures, which produce a posterior belief distribution over possible parameter values, seem naturally suited to the task of ABM calibration \citep{lamperti_agent-based_2018, platt_bayesian_2022, dyer_black-box_2024}. This is further reinforced by the fact that domain expert advice can be encoded into the Bayesian inference procedure through the specification of prior beliefs. However, given the complexity of systems being modelled, any ABM constructed to study a complex system will likely imperfectly capture the real-world data-generating process—that is, it will be \textit{misspecified} to some extent. Traditional Bayesian inference can be sensitive to such misspecification, motivating the use of more robust approaches \citep{cannon_investigating_2022, knoblauch_optimization-centric_2022}. These robust alternatives are collectively known as \emph{generalised Bayesian} methods, which modify standard Bayesian inference to better handle model misspecification. Additionally, the high computational cost of ABM simulation creates a need for sample-efficient calibration procedures that can reproduce empirical data with few simulations. These dual challenges—robustness to misspecification and computational efficiency—motivate our focus on differentiable ABMs, which can enable both generalised Bayesian inference methods and gradient-based optimization techniques.

While several frameworks for \emph{differentiating ABMs} have recently emerged \citep{andelfinger_differentiable_2021, chopra_differentiable_2023,dyer_gradient-assisted_2023}, no systematic analysis of their gradient estimation accuracy and reliability across diverse ABMs has been conducted.

\paragraph{Contribution} Motivated by this gap, we conduct a comprehensive evaluation of gradient estimation techniques for ABMs, comparing pathwise derivatives obtained through AD against finite difference approximations. Through detailed implementation walk-throughs, we provide a general recipe for building \emph{pathwise differentiable} ABMs. We demonstrate the effectiveness of the method proposed by implementing AD in three distinct models of relevance in the ABM literature: Axtell's model of firms \citep{axtell_emergence_1999}; Sugarscape \citep{epstein_growing_1996}; and an epidemiological agent-based SIR model \citep{epstein_modelling_2009}, each presenting unique challenges for gradient estimation. Our analysis demonstrates the conditions under which various gradient estimation techniques are reliable and identifies scenarios where they may fail, particularly when ABMs deviate significantly from mean-field approximations.

We also address the computational challenges of ABM calibration using AD. Specifically, we study the combination of gradient estimation for ABMs with generalised variational inference \citep{rezende_variational_2015,knoblauch_optimization-centric_2022, dyer_black-box_2024, quera-bofarull_bayesian_2023}, consolidating existing approaches into a robust calibration procedure that exploits gradient information. Empirical experiments demonstrate that this approach performs parameter inference for ABMs more efficiently than equivalent procedures that do not employ model gradients. Finally, we show how ABM gradients enable \emph{one-shot} sensitivity analysis. The gradient provides a first-order approximation of ABM sensitivity at given parameter values in a single simulation run, eliminating the need for many expensive simulations during sensitivity analysis.

\paragraph{Contents and Structure} \autoref{sec:abms} introduces ABMs using the Axtell's model of firms, Sugarscape, and an SIR epidemiological model as examples. \autoref{sec:autodiff} reviews AD and its adaptation for highly stochastic and discrete ABMs. \autoref{sec:diff_abms} provides explicit walk-throughs showing how each ABM can be modified to ensure pathwise differentiability. \autoref{sec:grad_val} benchmarks pathwise gradient estimation approaches against finite difference approximations, and demonstrates how gradient estimation enables rapid and interpretable sensitivity analysis. \autoref{sec:calib} reviews ABM calibration and its connection to simulation-based inference, empirically demonstrating how gradient estimation techniques can be combined with generalised variational inference to expedite ABM calibration. \autoref{sec:lit} reviews relevant literature, and \autoref{sec:discussion} concludes.

\section{Agent-Based Models (ABMs)}
\label{sec:abms}

Agent-based modelling spans a broad range of application domains, making it difficult to provide a single mathematical definition that captures all possible use cases. We therefore offer a general working definition sufficient for our purposes. To aid readers unfamiliar with agent-based modelling, we subsequently examine three seminal ABMs in detail: Axtell's model of firms; Sugarscape; and an epidemiological SIR model. These models will also serve as benchmarks for the gradient estimation techniques discussed later in the paper.

Generally speaking, an ABM is a stochastic map $f: \Theta \to \mathcal{X}$ implicitly defined through a synthetic population of agents interacting within an environment. Agents typically represent real-world entities that interact and make decisions. In economic ABMs, for instance, agents may represent firms, banks, or individual investors, depending on the granularity specified by the modeller. More broadly, what constitutes an agent depends heavily on the problem context and the modeller's objectives.

The parameter space $\Theta$ describes possible parametrisations of the ABM. In a epidemiological ABM, for instance, one parameter setting $\theta_{1} \in \Theta$ may correspond to a highly contagious virus, whilst another parameter setting $\theta_{2} \in \Theta$ may correspond to a virus that spreads at a slower rate. Beyond dictating the behaviour of the environment, parameters may also be responsible for adjusting agent behaviour. For example, in a financial ABM, adjusting the parameter value $\theta$ may alter the risk attitude of agents representing investment firms. 

The output space $\mathcal{X}$ tracks the internal states of each agent within the population through time, as they interact with each other and the environment. In an epidemiological ABM, the ABM output may track the infection status of each individual over a discrete time horizon. Likewise, in an opinion dynamics model, ABM output may track the political bias of each user within a social network through time. 

Most modellers focus on \emph{emergent} properties, which may be formulated as a function $\Phi: \mathcal{X} \to \mathbb{R}^{n}$, mapping outputs $x \in \mathcal{X}$ to an $n$-dimensional vector of summary statistics. For an epidemiological model, $\Phi(x)$ may correspond to a time series tracking the total number of infected individuals in the population, where $n$ corresponds to the time horizon simulated.

Agent interactions are traditionally specified through a set of heuristic rules, informed by a domain-expert, which depend on the parameters $\theta$. Several examples of such rules are given in Section \ref{sec:diff_abms}. More recent approaches include the use of reinforcement learning algorithms and large language models to specify agent behaviour and the dynamics of the environment. Whilst we focus on ABMs with handcrafted rule-sets in this work, we believe that the methods we propose would yield similar benefits for other ABMs as well \citep{chopra_limits_2024}.

Note that agent-based modelling shares some similarities with multi-agent reinforcement learning (MARL)~\citep{marl2025}. However, we emphasise that \emph{the goals of MARL and agent-based modelling are fundamentally different}. Typically, the goal of MARL is to train a small group of complex agents to maximise some (cooperative or non-cooperative) reward function. In contrast, ABMs are models aiming to reproduce and increase understanding of emergent phenomena in complex systems through the simulation of large-scale populations of simple agents.

In the subsections that follow, we review several classical ABMs, representative of the wider literature, in detail. These ABMs will also serve as the basis for our empirical experiments in Sections \ref{sec:grad_val} and \ref{sec:calib}. The familiar reader may skip to the end of each subsection, where a short summary of each model is provided, and refer back later as necessary.

\subsection{Axtell's Model of Firms}
\label{sec:amof}

Axtell's model of firms (AMOF) \citep{axtell_emergence_1999} simulates firms formation through agents optimizing work-leisure balance, reproducing key macroscopic phenomena including power-law firm size distributions and double-exponential log-growth rate distributions. The following subsection provides a formal definition of our AMOF variation used throughout this analysis.

AMOF consists of a population of $N$ utility-maximizing agents. Each agent $i$ has an individual preference between work and leisure characterized by a value $\theta_{i} \in [0, 1]$. We assume that each agent's preference value is independently sampled from a Beta distribution with parameters $\theta_{\alpha}$ and $\theta_{\beta}$. When $\theta_{i} \approx 0$, agent $i$ has a strong preference for leisure, while when $\theta_{i} \approx 1$, agent A friendship network models/captures the agents' social connections. 

To generate output, agents collaborate within firms over discrete time steps. On each time step $t$, each agent chooses their optimal effort level $e_{i}(t) \in [0, 1]$ as a decision variable to maximize their utility. We assume that each agent's initial effort level $e_{i}(0)$ is sampled independently from a Beta distribution with parameters $e_{\alpha}$ and $e_{\beta}$. The total effort level of a firm on time step $t$ is simply the sum of its agents' effort levels
\begin{equation}
E(f, t) = \sum_{i \in f} e_{i}(t).
\end{equation}

When the firm $f$ and time step $t$ are clear from context, we will refer to $E(f, t)$ by $E$ for ease of presentation. Likewise, we use $E_{-i}$ to denote the total firm effort level excluding the contribution of agent $i$. That is $E_{-i}(f, t) := E(f, t) - e_{i}(t)$. Finally, following set theoretic notation, we use $f \cup \{i\}$ to denote the firm formed when agent $i$ joins an existing firm $f$.

The effort level of a firm determines the output it produces. Specifically, the output of a firm $f$ is determined by the production function,
\begin{equation}
O(E, f) = a_{f}E + b_{f}E^2,
\end{equation}
where $a_{f}, b_{f} \in [0, 1]$ are firm-dependent positive parameters referred to as the production scale and increase in returns, respectively. Note that the production function is increasing in $E$, so firms with higher effort levels produce more output. The quadratic term promotes firm formation and cooperation: two agents produce more output by working together than by working alone, all else being equal. The production scale $a_{f}$ determines the linear effect of firm effort on output, while the increase in returns $b_{f}$ amplifies or dampens the benefits of cooperation within the firm. We assume that each firm's production scale and increase in returns are sampled independently from Beta distributions with parameters $a_{\alpha}, a_{\beta}$ and $b_{\alpha}, b_{\beta}$, respectively.

At the end of each time step, agents divide firm output equally amongst themselves and compute their utility according to a Cobb-Douglas preference model. Specifically, the utility attained by agent $i$ working at effort level $e$ in firm $f$ is given by
\begin{equation}
U_i(e, f) = \left(\frac{O(e + E_{-i}, f)}{|f|}\right)^{\theta_i} (1-e)^{1-\theta_i},
\end{equation}
where $|f|$ is number of agents belonging to firm $f$. The first multiplicative term represents the agent's share of firm output from their work, while the second term corresponds to leisure and decreases with effort level. The parameter $\theta_{i}$ controls the agent $i$'s preference for work versus leisure.

After evaluating their utility agents choose between (i) staying at their current firm, (ii) switching to an existing firm, and (iii) starting their own firm via the following procedure. Let $\mathcal{F}_{i}$ denote the set containing agent $i$'s firm, the firms of its friends, and the singleton firm containing only agent $i$. For each firm in $f \in \mathcal{F}_{i}$, agent $i$ solves the optimisation problem
\begin{equation*}
     U^{\star}_{i}(f) = \max_{e \in [0, 1]} U_{i}(e, f\cup\{i\}).
\end{equation*}
That is, agent $i$ evaluates the maximum utility it could achieve by working at each firm in $\mathcal{F}_{i}$ with optimally adjusted effort level. The agent then chooses to join the firm in $ \mathcal{F}_{i}$ that would maximize its utility and sets its effort level for the next time step accordingly. More formally, agent $i$ joins firm $f^{\star} \in \mathcal{F}_{i}$,
\begin{equation*}
    f^{\star} = \arg\max_{f \in \mathcal{F}_{i}} \: U_{i}^{\star}(f),
\end{equation*}
and sets their effort level $e_{i}(t+1)$ to $e^{\star}$,
\begin{equation*}
    e^{\star} = \arg\max_{e \in [0, 1]}U_{i}(e, f^{\star}  \cup \{i\}).
\end{equation*}
As demonstrated by \cite{axtell_emergence_1999}, $U^{\star}_{i}(f)$ has an analytical solution which is easy to evaluate. Thus, since $\mathcal{F}_{i}$ is finite, one may identify both $f^{\star}$ and $e^{\star}$ by iterating over a finite set of values.


\paragraph{Summary} AMOF consists of utility-maximizing agents who form and switch between firms to optimize their work-leisure balance. Each agent must solve a continuous optimization problem to determine their optimal effort level for each potential firm, then make a discrete choice by selecting the firm that maximizes their utility via an argmax operation. The model's non-differentiable elements include this argmax operation for firm selection, the subsequent discrete update of agent effort levels based on this choice, and the need for gradients to flow through the discrete updates to firm size $|f|$. The model parameters listed in Table \ref{tab:properties_axtell} represent the quantities with respect to which we will later compute gradients using automatic differentiation.

\begin{table}[h]
\centering
\begin{tabular}{llll}
\hline
Property & Level & Description & Sampled from \\
\hline
$\theta_{i}$ & Agent & Preferred trade-off between income and leisure & $\text{Beta} (\theta_{\alpha}, \theta_{\beta})$ \\
$e_{i}(0)$ & Agent  & Initial effort level & $\text{Beta}(e_{\alpha}, e_{\beta})$ \\
$a_{f}$ & Firm & Production scale & $\text{Beta}(a_{\alpha}, a_{\beta})$ \\
$b_{f}$ & Firm & Increase in return & $\text{Beta}(b_{\alpha}, b_{\beta})$ \\
\hline
\end{tabular}
\caption{Properties of agents and firms within AMOF. Note that our implementation of AMOF has eight parameters in total: $\theta_{\alpha}, \theta_{\beta}, e_{\alpha}, e_{\beta}, a_{\alpha}, a_{\beta}, b_{\alpha}, b_{\beta}$.}
\label{tab:properties_axtell}
\end{table}

\subsection{Sugarscape}
\label{sec:sugarscape}

Sugarscape \citep{epstein_growing_1996} is an agent-based model that studies artificial society dynamics by placing agents on a two-dimensional grid where they harvest sugar---a regenerating resource---to survive. Like Conway's Game of Life \citep{gardner_mathematical_1970} and Schelling's segregation model \citep{schelling_dynamic_1971}, agents move autonomously to maximize survival, yet generate complex emergent behaviors. Wealth distribution proves highly sensitive to individual behaviors, demonstrating how macroeconomic phenomena arise from collective actions. While extensions incorporating reproduction, multiple resources, or group affiliations can reproduce real-world phenomena like tribalism and trade \citep{epstein_growing_1996}, we focus on the basic version to illustrate fundamental challenges of applying automatic differentiation within agent-based modeling frameworks.

More formally, Sugarscape simulates a population of $N$ agents moving on a two-dimensional toroidal grid of $M \times M$ cells over $T$ discrete time steps. The grid is wrapped as a torus, meaning agents can move seamlessly from one edge to the opposite edge. Each agent $i$ possesses two time-invariant properties: vision radius $v_{i}$ and metabolic rate $m_{i}$. Each agent's metabolic rate is sampled independently from a Beta distribution with parameters $m_{\alpha}$ and $m_{\beta}$, while vision range is drawn from a categorical distribution $\text{Cat}(p)$ supported on a finite set $\mathcal{V} = \{v_{1}, \ldots, v_{k}\}$ of possible vision ranges. Additionally, each agent $i$ maintains sugar holdings $h_{i}(t)$ that evolve over time. Initial holdings $h_{i}(0)$ are sampled independently from a Beta distribution with parameters $w_{\alpha}$ and $w_{\beta}$.

At each time step's conclusion, agent $i$ must metabolize $m_{i}$ units of sugar from their holdings to survive. Prior to metabolism, agents may relocate to new grid positions and harvest available sugar. We denote the sugar contained in grid cell $(x, y)$ at time $t$ as $s(x, y, t)$. To select a destination, each agent $i$ observes unoccupied cells within a Von Neumann neighbourhood of radius $v_{i}$ centred at their current location $(x_{i}(t), y_{i}(t))$. Agents move to the visible cell containing the most sugar,
\begin{equation}
    \label{eq:sugarscape_argmax}
    (x_{i}(t+1), y_{i}(t+1)) = \argmax_{(x, y) \in \mathcal{N}_{i}(t)} s(x, y, t),
\end{equation}
where $\mathcal{N}_{i}(t)$ denotes the unoccupied cells observed by agent $i$ at time $t$. The vision radius $v_{i}$ captures an agent's foraging effectiveness. To prevent collisions, agents move in randomized order each time step. After relocating, agent $i$ updates their holdings as
\begin{equation*}
    h_{i}(t+1) = h_{i}(t) + s(x_{i}(t+1), y_{i}(t+1), t) - m_{i}.
\end{equation*}
Agents harvest sugar from their new location and metabolize a portion of their holdings. If $h_{i}(t+1) \leq 0$, agent $i$ dies from insufficient resources. 

Sugar reserves on each grid cell regenerate according to
\begin{equation}
\label{eq:regenrule}
    s(x, y, t+1) = \min \{s(x, y, t) \cdot [1 - o(x, y, t)] + r, c(x, y)\},
\end{equation}
where $o(x, y, t)$ indicates whether cell $(x, y)$ was harvested at time $t$, $r$ represents the global regeneration rate, and $c(x, y)$ denotes cell capacity. Unharvested cells regenerate $r$ units of sugar, capped at their maximum capacity $c(x, y)$.

\begin{figure}[h]
    \centering
    \includegraphics[width=0.8\linewidth]{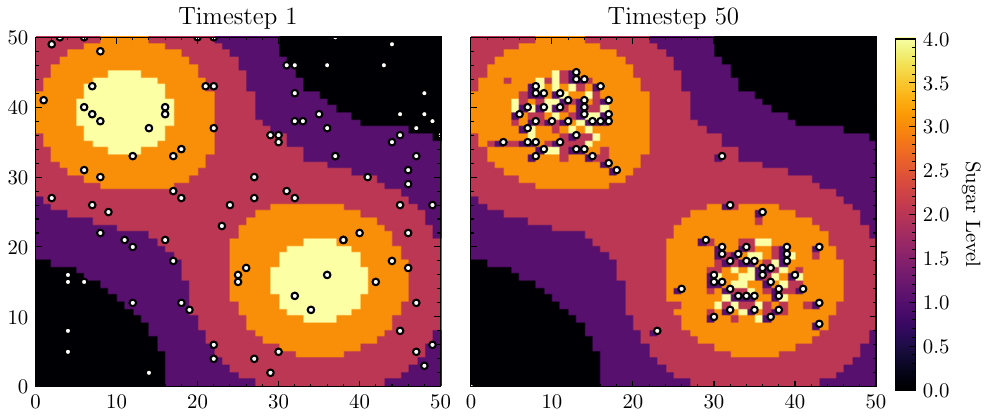}
    \caption{The first and final time steps of a Sugarscape simulation. Warmer colours indicate grid cells with higher sugar reserves. In this case, the landscape is characterised by two high capacity sugar peaks. Agents, represented by white dots, are initially dispersed, but accumulate at each peak towards the end of the simulation.}
    \label{fig:sugarscape}
\end{figure}

The first and last time steps of a Sugarscape simulation is depicted in Figure \ref{fig:sugarscape}. In this simulation, the landscape consists of two high capacity sugar peaks. Whilst agents are initially dispersed, they naturally cluster around each high capacity peak towards the end of the simulation.

\paragraph{Summary} Sugarscape consists of agents exploring a two-dimensional grid to harvest sugar for survival. Each agent must metabolize a fixed amount of sugar per time step or die, creating a discrete life-or-death threshold. Agents move by selecting the unoccupied cell with maximum sugar within their vision range—a discrete optimization problem involving both the argmax operation and occupancy constraints. The model's dynamics are thus fundamentally non-differentiable due to these discrete decision rules and binary survival outcomes. The model parameters listed in \autoref{tab:properties_sugar} represent the quantities with respect to which we will later compute gradients using automatic differentiation.

\begin{table}[h]
\centering
\begin{tabular}{llll}
\hline
Property & Level & Description & Sampled from  \\
\hline
$m_{i}$ & Agent & Units of sugar required to metabolise  & $\text{Beta}(m_{\alpha}, m_{\beta})$ \\
$v_{i}$ & Agent & Vision radius & $\text{Cat}(p)$   \\
$h_{i}(0)$ & Agent & Initial sugar holdings & $\text{Beta}(w_{\alpha}, w_{\beta})$ \\
$c(x, y)$ & Cell & Sugar capacity  & Fixed per experiment \\
$r$ & Global & Global sugar regeneration rate & Fixed per experiment  \\
\hline
\end{tabular}
\caption{Properties of agents, cells and the global environment in Sugarscape. Note that the regeneration rate of sugar is not unique to each cell, but instead fixed and uniform across the landscape. Our implementation has $5$ parameters in total: $m_{\alpha}, m_{\beta}, p, w_{\alpha}$ and $w_{\beta}$.}
\label{tab:properties_sugar}
\end{table}

\subsection{An SIR Agent-Based Epidemiological Model}
\label{sec:sir_intro}

Finally, we consider a network-based epidemiological model where agents occupy Susceptible, Infected, or Recovered compartments, with disease transmission occurring through contact network interactions. While compartmental models have been central to epidemiology since the early 20th century \citep{kermack_contribution_1927}, agent-based epidemiological models proved essential during COVID-19 for policy testing \citep{hinch_openabm-covid19agent-based_2021, kerr_covasim_2021, aylett-bullock_june_2021}, modeling individuals explicitly rather than using aggregate differential equations. This network-based compartmental model captures the fundamental challenges of applying differentiable methodologies to epidemiological models with explicit individual representation.

The model consists of $N$ agents, each with an individual state $s_{i}(t) \in \{\text{S}, \text{I}, \text{R}\}$ that evolves through time and corresponds to one of three compartments. Agents in state S are \textbf{S}usceptible to infection, agents in state I are currently \textbf{I}nfected, and agents in state R have \textbf{R}ecovered and gained immunity. Each agent has a probability $I_{0} \in [0, 1]$ of being initially infected ($s_{i}(0) = \text{I}$), and is otherwise initialized as susceptible ($s_{i}(0) = \text{S}$). 

Agents are connected by a contact network $\mathcal{G}$, which remains fixed throughout the simulation. The contact network represents potential transmission pathways between agents based on their social interactions. For example, agents who work together or live in the same household are likely to be connected by an edge in $\mathcal{G}$, while agents with no social contact are not connected.

Disease transmission occurs through direct contact between connected agents over discrete time steps. At each time step, every agent potentially interacts with all of their network neighbours. The force of infection $\lambda_{i}(t)$ for agent $i$ at time $t$ is given by
\begin{equation}
   \lambda_{i}(t) = 
   \begin{cases}
       \beta \cdot \frac{1}{|\mathcal{N}(i)|}\sum_{j \in \mathcal{N}(i)} \mathbbm{1}_{\{s_{j}(t) = \text{I}\}} \quad & \text{if } \: s_{i}(t) = \text{S}, \\
       0 \quad & \text{otherwise},
   \end{cases}
   \label{eq:sir_standard_dynamics}
\end{equation}
where $\mathcal{N}(i)$ denotes agent $i$'s immediate neighbours in $\mathcal{G}$. The force of infection is zero for infected and recovered agents, as they cannot be reinfected. For susceptible agents, the force of infection equals the transmission rate $\beta \in \mathbb R^+ $ multiplied by the fraction of neighbours who are currently infected. The parameter $\beta$ captures the infectiousness of the pathogen. Note that infection risk depends on the fraction of infected neighbours, not the absolute number.

The probability that an agent $i$ becomes infected during the current time step is $p_{i}(t) = 1 - \exp(-\lambda_{i}(t)\Delta t)$, where $\Delta t$ is the discrete time step length. Intuitively, an agent is more likely to become infected during a longer time step. If a susceptible agent becomes infected, its state updates for the next time step ($s_{i}(t+1) = \text{I}$). 

Simultaneously, each infected agent $i$ recovers with probability $q_{i} = 1 - \exp(-\gamma \Delta t)$, where $\gamma \in \mathbb R^+$ is the recovery rate parameter. The recovery rate models how quickly the population clears the infection once infected. Higher values of $\gamma$ correspond to faster recovery. When an infected agent recovers, its state updates to recovered ($s_{i}(t+1) = \text{R}$), and it remains in this state permanently, representing acquired immunity.

Beyond the baseline dynamics described above, we consider two policy interventions that modify agent behaviour: \textit{quarantine} and \textit{social distancing} policies. 

A quarantine policy $Q$ is characterized by a start time $Q_{\text{start}}$, an end time $Q_{\text{end}}$, and a compliance probability $p_{Q}$. The compliance probability $p_{Q}$ captures the likelihood that infected agents will follow quarantine guidelines when implemented. Formally, during any time step $t \in \{Q_{\text{start}}, \dots, Q_{\text{end}}\}$, each infected agent $i$ complies with quarantine rules with probability $p_{Q}$. We denote by $Q_{i}(t)$ the binary indicator of whether agent $i$ is quarantining at time step $t$, where $Q_{i}(t) = 1$ indicates compliance and $Q_{i}(t) = 0$ indicates non-compliance. We define $\mathcal{N}(i ,t) := \{j \in \mathcal{N}(i) \: : \: Q_{j}(t) = 0\}$ as the set of agent $i$'s neighbours who are not quarantining at time step $t$.

Under quarantine intervention, disease transmission is restricted to interactions between non-quarantining agents only. For time steps $t \in \{Q_{\text{start}}, \dots, Q_{\text{end}}\}$, the force of infection becomes
\begin{equation}
    \lambda_{i}(t) = 
    \begin{cases}
        \beta \cdot \frac{1}{|\mathcal N(i, t)|} \sum_{j \in \mathcal{N}(i, t)} \mathbbm{1}_{\{s_{j}(t) = \text{I}\}} \quad & \text{if } \: s_{i}(t) = \text{S} \text{ and }  Q_{i}(t) = 0,\\
        0 \quad & \text{otherwise}.
    \end{cases}
    \label{eq:sir_q_dynamics}
\end{equation}
This formulation ensures that susceptible agents can only become infected if they are not quarantining themselves, and they can only be exposed to infected neighbours who are also not quarantining.

The second policy intervention is social distancing. A social distancing policy $D$ is similarly characterized by a start time $D_{\text{start}}$, an end time $D_{\text{end}}$, and a transmission reduction factor $\alpha_{D} \in [0, 1)$. During social distancing periods, all agents modify their interaction behaviour to reduce transmission risk through measures such as mask-wearing, maintaining physical distance, or reducing contact duration. The reduction factor $\alpha_{D}$ quantifies the effectiveness of these precautionary measures, where smaller values of $\alpha_{D}$ correspond to greater reductions in transmission risk.

During social distancing intervention periods $t \in \{D_{\text{start}}, \dots, D_{\text{end}}\}$, the force of infection is modified as:
\begin{equation}
    \lambda_{i}(t) = 
    \begin{cases}
        \alpha_{D} \cdot \beta \cdot \frac{1}{|\mathcal N(i)|}\sum_{j \in \mathcal{N}(i)} \mathbbm{1}_{\{s_{j}(t) = \text{I}\}} \quad & \text{if } \: s_{i}(t) = \text{S}, \\
        0 \quad & \text{otherwise}.
    \end{cases}
    \label{eq:sir_sd_dynamics}
\end{equation}
In effect, social distancing scales down the effective transmission rate from $\beta$ to $\alpha_{D} \cdot \beta$, uniformly reducing infection probability across all agent interactions while maintaining the same contact network structure.

\paragraph{Summary} We study a network-based SIR agent-based model where $N$ agents with discrete states $s_{i}(t) \in \{\text{S}, \text{I}, \text{R}\}$ interact over a fixed contact network $\mathcal{G}$. Disease transmission occurs stochastically based on infected neighbours and parameters $\beta$ (transmission rate), $\gamma$ (recovery rate), and $I_0$ (initial infection probability). We consider two policy interventions: quarantine (with compliance probability $p_Q$ and timings $Q_\mathrm{start}$, $Q_\mathrm{end}$) and social distancing (with reduction factor $\alpha_D$ and timings $D_\mathrm{start}$, $D_\mathrm{end}$). This model presents key challenges for differentiable optimization: discrete agent states and stochastic state transitions introduce non-smooth dynamics, while policy activation timings create discontinuous changes in model behaviour that are difficult to differentiate through. Additionally, quarantine compliance decisions and initial infection probabilities involve discrete random sampling that breaks gradient flow. Table \ref{tab:properties_sir} summarizes the 9 model parameters with respect to which we seek to compute gradients for policy optimization and parameter inference.

\begin{table}[h]
\centering
\begin{tabular}{llll}
\hline
Property & Level & Description  \\
\hline
$\beta$ & Global & Transmission rate  \\
$\gamma$ & Global & Recovery rate   \\
$I_{0}$ & Global & Initial infection probability \\
$Q_{\text{start}}$ & Quarantine policy & Quarantine start time \\
$Q_{\text{end}}$ & Quarantine policy & Quarantine end time  \\
$p_{Q}$ & Quarantine policy & Quarantine obedience probability  \\
$D_{\text{start}}$ & Social distancing policy & Social distancing start time \\
$D_{\text{end}}$ & Social distancing policy & Social distancing end time  \\
$\alpha_{D}$ & Social distancing policy & Reduction rate \\
\hline
\end{tabular}
\caption{Properties of the network-based SIR model presented in Section \ref{sec:sir_intro} alongside the properties of quarantine and social distancing interventions. In total, there are 9 parameters. }
\label{tab:properties_sir}
\end{table}


\section{Automatic Differentiation}
\label{sec:autodiff}

Before examining how each of the ABMs introduced in the previous section may be made differentiable, we provide a brief review of Automatic Differentiation (AD). AD encompasses a family of techniques for computing exact partial derivatives of functions defined by computer programs. To appreciate the advantages of AD, we first examine classical numerical approaches for computing derivatives using finite differences.

Finite-difference (FD) methods directly apply the limit definition of the derivative for numerical approximation. For any smooth function $f: \mathbb{R}^{m} \to \mathbb{R}^{n}$ and sufficiently small $\epsilon > 0$, the partial derivative can be approximated as
\begin{equation}
    \label{eq:fd}
    \frac{\partial f_{j}}{\partial x_{i}} \approx \frac{f(x + \epsilon e_{i}) - f(x - \epsilon e_{i})}{2\epsilon},
\end{equation}
where $e_{1}, \dots, e_{n}$ are the canonical basis vectors. 


FD approaches have fundamental limitations: they require careful step size selection (balancing truncation vs. floating-point errors) and scale poorly with dimensionality, requiring $2m$ function evaluations for gradients of $f: \mathbb{R}^{m} \to \mathbb{R}$.


AD overcomes these issues by decomposing programs into elementary operations with known derivatives and applying the chain rule systematically. Programs are represented as computational graphs---directed acyclic graphs where nodes are operations and edges represent data flow. As noted by \cite{laue_equivalence_2022}, AD is algorithmically equivalent to symbolic differentiation.

Program evaluation (the \emph{primal pass}) traverses the graph from root to leaf nodes. The \emph{tangent pass} computes derivatives via the chain rule and can proceed in two directions:
\begin{enumerate}
    \item \emph{Forward-mode differentiation}: Performs the tangent pass simultaneously with the primal pass by traversing forward from root to leaf nodes. Typically implemented using dual numbers that pair each value with its derivative. Computational complexity scales linearly with the number of inputs and is constant with respect to outputs.
    
    \item \emph{Reverse-mode differentiation}: First performs the primal pass, then traverses backward from leaf to root nodes applying the chain rule. Complexity remains constant with respect to inputs and scales linearly with outputs.
\end{enumerate}

Forward-mode AD scales with inputs while reverse-mode scales with outputs. Since deep learning models have millions of parameters but single loss outputs, reverse-mode AD dominates machine learning applications. However, forward-mode AD has a key advantage: it performs primal and tangent passes simultaneously, discarding intermediate values immediately, while reverse-mode must store all intermediate values from the primal pass for later use. This makes forward-mode significantly more memory-efficient, which has important implications for ABMs as discussed in \autoref{sec:sa}.

\subsection{Automatic Differentiation and Agent-Based Models}
Recall that an ABM may be interpreted as a stochastic map $f: \Theta \to \mathcal{X}$, from a parameter space $\Theta$ to an output space $\mathcal{X}$ which is implicitly defined through agents interacting within an environment. Generally speaking, one is interested in the average emergent behaviour of the model rather than the specific states of individual agents. For instance, in an epidemiological ABM, researchers typically focus on the expected number of infections over time or the average duration of an epidemic, rather than attempting to track whether any particular agent becomes infected. More formally, this interest in aggregate behaviour can be expressed as
\begin{equation}
    \label{eq:avgemerge}
    \mathbb{E}_{x \sim p(x\mid\theta)}\left[\Phi(x)\right],
\end{equation}
where $p(x\mid \theta)$ is the likelihood function associated with the stochastic map $f$, and $\Phi: \mathcal{X} \to \mathbb{R}^{n}$ maps model outputs to the associated emergent properties they describe. By computing derivatives of this expectation, we can investigate how the average emergent properties of a system change with respect to model parameters. However, applying AD to \autoref{eq:avgemerge} presents several challenges.

First, note that \autoref{eq:avgemerge} includes an expectation over a probability distribution that explicitly depends on $\theta$. The reparametrisation trick, popularised by \cite{kingma_auto-encoding_2022}, addresses this issue by designing a function $g$ and distribution $q$ independent of $\theta$ such that
\begin{equation}
    \epsilon \sim q(\cdot) \implies g(\epsilon ; \theta) \sim p(\cdot\mid \theta).
\end{equation}
In other words, sampling from the fixed distribution $q$ and applying the function $g(\cdot, \theta)$ is equivalent to sampling directly from the distribution $p(\cdot \mid \theta)$. If $g$ is sufficiently well-behaved (in the sense that the operations of differentiation and integration may be exchanged), then
\begin{align*}
    \nabla_{\theta}\mathbb{E}_{x \sim p(x\mid\theta)}[\Phi(x)]
    &= \nabla_{\theta}\mathbb{E}_{\epsilon \sim q(\epsilon)}[\Phi(g(\epsilon; \theta))] \\
    &= \mathbb{E}_{\epsilon \sim q(\epsilon)}[\nabla_{\theta}\Phi(g(\epsilon; \theta))] \\
    &\approx \frac{1}{m}\sum^{m}_{i=1}\nabla_{\theta}\Phi(g(\epsilon_{i}; \theta)), \quad \epsilon_{i} \sim q(\cdot),
\end{align*}
where the law of the unconscious statistician (LOTUS) was applied in the first equality and the Leibniz rule for differentiation under the integral sign was applied in the second. Put differently, this approach allows the derivative to be estimated using Monte Carlo approximation. Observe that any practical implementation of an ABM takes the form of a stochastic program describing the reparametrisation $g$. Thus, in principle, AD may be applied to the resulting computational graph to estimate derivatives of the form $\nabla_{\theta}\Phi(g(\epsilon_{i}, \theta))$.

However, for AD to be applied, the constituent nodes comprising the computational graph associated with $g$ must correspond to differentiable operations. Unfortunately, in many ABMs, agents and the environment often transition between discrete states. For example, in the SIR model, agents transition between susceptibility and infection. Likewise, agents in Sugarscape die once their sugar reserves are exhausted. Such transitions between discrete states are handled programmatically via \emph{control flow statements}, whose predicates are often dependent on model parameters. Differentiating through control statements presents a significant challenge especially when different branches entail sharp changes in agent behaviour.

Moreover, agents are often tasked with optimising over a \emph{discrete} set of actions. For instance, in AMOF, agents select the firm that maximises their personal utility from those available. Likewise, in Sugarscape, each agent identifies the unoccupied cell with the largest sugar reserve within its vision radius before moving. The discreteness of the action space in such settings means that the classical notion of the derivative is no longer well-defined, and more general approaches are required.

Lastly, ABMs frequently employ \emph{discrete randomness} that cannot be readily reparameterised. For example, in the SIR model, an agent's chance of infection is modelled as a Bernoulli random variable whose parameter is dependent on the agent's number of infected neighbours. There is no natural reparametrisation of this random sampling procedure that is smooth and permits the application of Leibniz rule.

These challenges --- \emph{control flow statements}, \emph{optimisation over finite sets}, and \emph{discrete sampling procedures} --- cannot be reparameterised in a differentiable manner during forward simulation of ABMs. In what follows, we discuss a range of surrogate gradient methods that mitigate these issues. In short, surrogate gradient methods replace non-differentiable operations, such as control flow statements, with differentiable surrogates during the tangent pass, so that a (slightly biased) estimate of the derivative can be obtained via Monte Carlo sampling. Crucially, surrogate gradient methods leave the primal pass unchanged, ensuring that the ABM's output remains unaltered. After discussing surrogate gradients, we will introduce an alternative approach for differentiating through discrete sampling procedures that is unbiased. 

\subsection{Surrogate Gradient Methods}

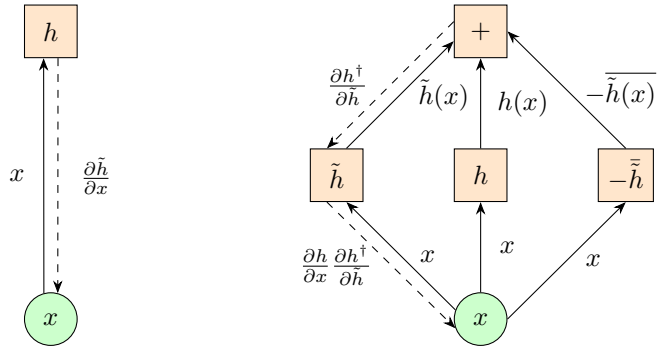
\begin{figure}[h]
    \centering
    \begin{tikzpicture}[
    node distance=1.2cm,
    every node/.style={circle, draw, minimum size=0.7cm, font=\small},
    var/.style={fill=green!20},
    operation/.style={fill=orange!20, rectangle},
    >=Stealth
]

\node[operation] (h) {$h$};
\node[operation, above =of h] (add) {$+$};
\node[operation, left =of h] (th) {$\tilde{h}$};
\node[operation, right =of h] (bth) {$-\bar{\tilde{h}}$};
\node[var, below =of h] (x) {$x$};
\node[operation, left=5 of add] (h_simple) {$h$};
\node[var, left =5 of x] (x_simple) {$x$};

\draw[->, dashed] (h_simple.285) -- (x_simple.75) node[midway, right, draw=none] {$\frac{\partial \tilde{h}}{\partial x}$};
\draw[->] (x_simple.105) -- (h_simple.255) node[midway, left, draw=none] {$x$};

\draw[->, dashed] (th.250) -- (x.200) node[midway, below, left, draw=none] {$\frac{\partial h}{\partial x}\frac{\partial h^{\dagger}}{\partial \tilde{h}}$};
\draw[->] (x.160) -- (th.290) node[midway, right, draw=none] {$x$};
\draw[->] (x) -- (h) node[midway, right, draw=none] {$x$};
\draw[->] (x.0) -- (bth.270) node[midway, below, right, draw=none] {$x$};
\draw[->] (h) -- (add) node[midway, below, right, draw=none] {$h(x)$};
\draw[->] (bth.90) -- (add.360) node[midway, below, right, draw=none] {$-\overline{\tilde{h}(x)}$};
\draw[->] (th.70) -- (add.200) node[midway, right, draw=none] {$\tilde{h}(x)$};
\draw[->, dashed]  (add.160) -- (th.110) node[midway, left, draw=none] {$\frac{\partial h^{\dagger}}{\partial \tilde{h}}$};
\end{tikzpicture}
    \caption{Two methods for implementing the surrogate gradient method. In the left graph, the tangent pass is edited directly, whilst in the right graph function calls to $h$ are replaced by function calls to $h^{\dagger}$.}
    \label{fig:comp_graphs}
\end{figure}

As mentioned above, the surrogate gradient method replaces a non-differentiable function $h$ with a smooth approximation $\tilde{h}$ during the tangent pass. An implementation of the surrogate gradient method that directly manipulates the tangent pass is depicted in Figure \ref{fig:comp_graphs} (left). Whilst compact, editing the computational graph in this manner is often complicated from a programming perspective since popular AD packages, such as \texttt{PyTorch} and \texttt{Tensorflow}, automatically perform the tangent pass under the hood. Instead it is common to implement the surrogate gradient method by replacing program calls to $h$ with calls to the function
\begin{equation*}
    h^\dagger(x) = h(x) + (\tilde{h}(x) - \overline{\tilde{h}(x)}),
\end{equation*}
where the bar notation requires explanation. For any function $a$, we define $\bar{a}(x)=a(x)$ for all possible inputs $x$. In other words, $\bar{a}$ is equivalent to $a$ from the primal perspective. During the tangent pass, however, all paths containing computational nodes corresponding to $\bar{a}$ are ignored. This is mathematically equivalent to setting $\frac{\partial \bar{a}}{\partial x} = 0$. Many AD packages support such an operation. For example, in \texttt{PyTorch}, $\bar{a}$ may be implemented by calling \texttt{a(x).detach()}, whilst in \texttt{Tensorflow}, $\bar{a}$ may be implemented by calling \texttt{tf.stop\_gradient(a(x))}. 

The computational graph corresponding to $h^{\dagger}(x)$ is depicted in Figure \ref{fig:comp_graphs} (right). It is straightforward to verify that $h^{\dagger}(x) = h(x)$, leaving the primal pass unaltered. During the tangent pass, all paths containing either $h$ or $\overline{\tilde{h}}$ are ignored, meaning that the chain rule is only applied along paths containing $\tilde{h}$.

Choosing an appropriate surrogate function may depend on the function being replaced during the tangent pass. We want to select surrogates that closely match the original function they are replacing in order to reduce the bias introduced during the tangent pass. However, we also want the surrogate function to yield informative gradients that do not vanish or explode during training.

\subsection{Differentiating Through Control Flow}
\label{sec:if_statements}

The methods presented in this section are largely inspired by the comprehensive review of \cite{blondel_elements_2024}. For a more detailed treatment of differentiable programming techniques, we refer readers to that work.

Control flow structures are fundamental components of discrete simulation models such as ABMs, enabling the expression of complex behaviours through conditionals, branches, and loops. However, these structures present significant challenges for AD when predicates depend on learnable parameters. We begin our investigation with the simplest control flow structure: the if-else statement,
\begin{equation}
    \text{ifelse}(\pi, v_1, v_0) = \begin{cases}
            v_1 & \text{if } \pi = 1, \\
            v_0 & \text{if } \pi = 0,
\end{cases}
\end{equation}
where $\pi$ denotes the boolean predicate that determines branch selection, and $v_1$, $v_0$ represent the operations executed in each respective branch. 

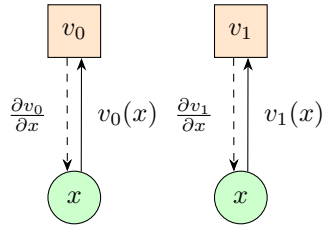
\begin{figure}[h]
\centering
\begin{tikzpicture}[
    node distance=1.5cm,
    every node/.style={circle, draw, minimum size=0.7cm, font=\small},
    var/.style={fill=green!20},
    operation/.style={fill=orange!20, rectangle},
    >=Stealth
]

\node[var] (x1) {$x$};
\node[var, right =of x1] (x2) {$x$};
\node[operation, above =of x1] (v0) {$v_{0}$};
\node[operation, above =of x2] (v1) {$v_{1}$};

\draw[->] (x1.75) -- (v0.285) node[midway, below, right, draw=none] {$v_{0}(x)$};
\draw[->, dashed] (v0.255) -- (x1.105) node[midway, below, left, draw=none] {$\frac{\partial v_{0}}{\partial x}$};

\draw[->] (x2.75) -- (v1.285) node[midway, below, right, draw=none] {$v_{1}(x)$};
\draw[->, dashed] (v1.255) -- (x2.105) node[midway, below, left, draw=none] {$\frac{\partial v_{1}}{\partial x}$};
\end{tikzpicture}
\caption{\label{fig:ifelse} Computational graphs generated by conventional AD engines for if-else statements. The left graph corresponds to $\pi = 0$ (executing $v_0$) and the right graph corresponds to $\pi = 1$ (executing $v_1$). Only the executed branch appears in the computational graph.}
\end{figure}

\paragraph{Parameter-independent case} When the predicate $\pi$ does not depend on differentiable parameters, most AD engines handle if-else statements without issue. This is because AD engines typically construct computational graphs dynamically during the forward pass. When an if-else statement is encountered, the predicate is evaluated and only the selected branch is executed and added to the graph. Without loss of generality, assume branch $v_0$ is selected; then $v_1$ is never evaluated, resulting in the computational graph shown in Figure \ref{fig:ifelse}.

While it may seem problematic that the unexecuted branch $v_1$ never contributes to gradient computation, this behaviour is mathematically correct. An if-else statement can be equivalently expressed as a masking operation,
\begin{equation}
\text{ifelse}(\pi, v_1, v_0) = \pi v_1 + (1-\pi)v_0.
\label{eq:maskif}
\end{equation}
The partial derivatives with respect to each branch are
\begin{equation}
\frac{\partial}{\partial v_0}\text{ifelse}(\pi, v_1, v_0) = (1-\pi)I, \quad \frac{\partial}{\partial v_1}\text{ifelse}(\pi, v_1, v_0) = \pi I,
\end{equation}
where $I$ denotes the identity matrix with dimensions matching the respective branch outputs. Since we executed branch $v_0$, we have $\pi = 0$, making the derivative with respect to $v_1$ exactly zero. Therefore, the unexecuted branch contributes no gradient information, justifying the AD engine's behaviour of ignoring it entirely.

\paragraph{The parameter-dependent case} The situation becomes significantly more complex when the predicate $\pi$ depends on differentiable parameters. Since $\pi$ is inherently binary, it introduces non-smoothness that renders $\text{ifelse}(\pi, v_1, v_0)$ generally non-differentiable with respect to those parameters.

For clarity, consider the case where $\pi$ takes the form of a one-sided comparison operator,
\begin{equation}
\pi = \mathrm{step}(x) = \begin{cases}
1 & \text{if } x \geq 0,\\
0 & \text{if } x < 0,
\end{cases}
\end{equation}
where $x \in \mathbb{R}$ is a scalar input dependent on learnable parameters. The analysis that follows applies analogously to predicates defined via two-sided comparison operators.

The fundamental challenge arises from the discontinuity at $x = 0$, where infinitesimal parameter changes can cause discrete branch switches during the forward pass. To address this, we replace the non-differentiable $\text{step}(x): \mathbb{R} \to {0, 1}$ with a smooth surrogate $s(x): \mathbb{R} \to [0, 1]$ that eliminates the discontinuity. Common choices include:
\begin{enumerate}
\item The Gaussian cumulative distribution function (CDF),
\begin{equation}
\psi(x) = \frac{1}{2}\left[1 + \mathrm{erf}\left(\frac{x}{\sqrt{2}\sigma}\right)\right],
\end{equation}
where $\mathrm{erf}$ is the error function and $\sigma$ controls the transition sharpness.
\item The sigmoid function,
\begin{equation}
S(x) = \frac{1}{1+\exp{(-kx)}},
\end{equation}
where $k$ determines the steepness of the transition.
\item A piecewise linear function,
\begin{equation}
\mathrm{pw}(x) = \begin{cases}
0 & \text{if } x \leq -a, \\
\frac{x + a}{b + a} & \text{if } -a < x < b,\\
1 & \text{if } x \geq b,
\end{cases}
\end{equation}
where $a$ and $b$ define the transition region boundaries.
\end{enumerate}

The choice of smoothing function and its hyperparameters involves a critical trade-off. The gradient should be sufficiently informative when branches are about to switch (providing learning signals for important decisions), but not so steep as to cause gradient explosion. Conversely, gradients should not vanish too rapidly away from the decision boundary, as this would require $x$ to be extremely close to zero for the alternate branch to contribute meaningful gradients. This balance depends heavily on how important branch switching is for the emergent behaviours of interest in the specific ABM.
A practical guideline is to match the function's variability range to the expected parameter variation within the model. For instance, choosing too large a $\sigma$ for Gaussian smoothing causes slow variation and weak gradients, while too small a $\sigma$ leads to rapid gradient vanishing away from $x = 0$.

\paragraph{Implementation via masking operations} Given a smooth step function $s(x)$, we implement differentiable if-else statements by replacing the predicate $\pi$ with $s(x)$ in the masking formulation,
\begin{equation}
\text{ifelse}(s(x), v_1, v_0) = s(x) v_1 + (1-s(x))v_0.
\label{eq:maskifsmooth}
\end{equation}
Unlike the discrete case where $\pi$ and $(1-\pi)$ are mutually exclusive, $s(x)$ and $(1-s(x))$ can be simultaneously positive. This requires executing both branches during the forward pass to compute complete gradients, contrasting with standard AD engines that execute only the selected branch. This issue is circumvented by implementing if-else statements directly as masking operations (\autoref{eq:maskifsmooth}) rather than using control flow keywords like \texttt{if} and \texttt{else}.

\begin{figure}[h]
    \centering
    \begin{subfigure}[b]{0.5\textwidth}
        \centering
        \includegraphics[width=\textwidth]{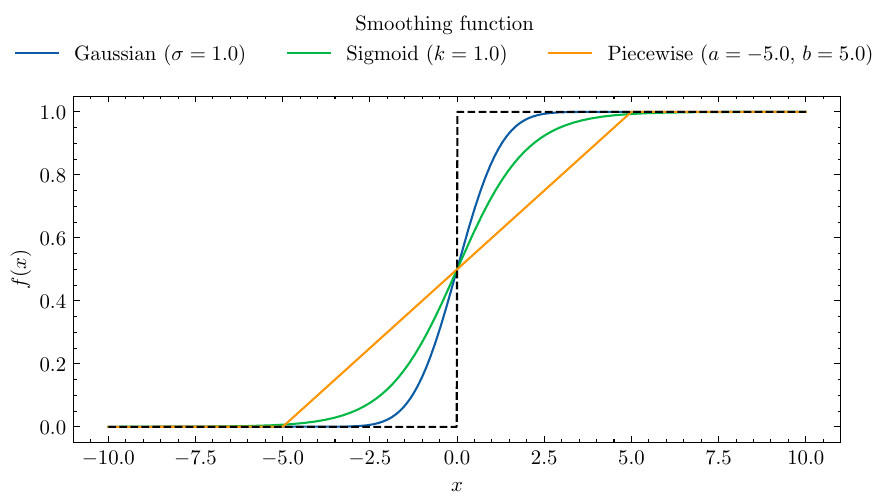}
        \caption{Different smoothing kernels for approximating the step function, enabling differentiable control flow.}
        \label{fig:if_smooth}
    \end{subfigure}
    \hfill
    \begin{subfigure}[b]{0.45\textwidth}
        \centering
        \begin{tikzpicture}[
    node distance={0.5cm and 0.5cm},
    var/.style={fill=green!20, circle, draw},
    operation/.style={fill=orange!20, rectangle, draw, minimum size=0.7cm},
    >=Stealth,
    edge from parent/.style={<-, draw},
    level 1/.style={level distance=1.2cm,sibling distance=1.5cm},
    level 2/.style={level distance=1.2cm,sibling distance=1cm},
    level 3/.style={level distance=1.2cm,sibling distance=1cm},
]

\node(0)[operation, draw]{$+$}
child{node(1)[operation]{$\times$}
    child{node(5)[operation] {$v_{0}$}}
    child{node(6) [operation] {$-$}
        child{node(7) [operation] {$1$}}
        child{node(8) [operation] {$\pi$}
            child{node(9) [var] {$x$}}}
    }
}
child{node(2)[operation]{$\times$}
    child[missing] {node(11)}
    child{node(10) [operation] {$v_{1}$}}
};

\draw[solid,bend left=60, ->](9.180) -| (5.270);
\draw[solid,bend right=60, ->](9.0) -| (10.270);
\draw[solid, ->](8.70) -> (2.240);

\end{tikzpicture}
        \caption{Computational graph for if-else statements implemented as masking operations (\autoref{eq:maskifsmooth}). Unlike \autoref{fig:ifelse}, both execution branches are included in the graph, enabling full gradient computation.}
        \label{fig:surrifelse}
    \end{subfigure}
    \hfill
\end{figure}

This approach incurs higher computational costs than non-differentiable control flow, as both branches must be evaluated. To prevent combinatorial explosion, it is crucial to avoid heavily nested control structures. Fortunately, ABMs are well-suited to this constraint: program execution is distributed across many agents employing relatively simple decision-making processes with shallow control flows, making differentiable control flow structures particularly amenable to agent-based modelling.

\subsubsection{Multi-Branch Structures}
\label{sec:multi_branch}

The masking approach extends naturally to multi-branch control flow structures. While if-else statements involve binary predicates taking values in $\{0, 1\}$, $K$-branch structures require predicates that can select among $K$ alternatives. Rather than using integer-valued predicates ${1, \ldots, K}$ (which are incompatible with the masking approach), we employ one-hot vector representations.

Specifically, the predicate is represented as $\pi \in {e_{1}, \ldots, e_{K}} \subset \Delta_{K-1}$, where $e_i$ denotes the $i$-th standard basis vector and $\Delta_{K-1}$ is the $(K-1)$-simplex. The non-zero element of $\pi$ indicates which branch should be executed. A $K$-branch control flow statement can then be expressed as

\begin{equation}
    \label{eq:cond_expanded}
    \text{cond}(\pi, v_1,\ldots,v_K) = \begin{cases}
        v_1 & \text{if } \pi = e_1 \\
        \vdots & \vdots \\
        v_K & \text{if } \pi = e_K,
    \end{cases}
\end{equation}
where $v_{1}, \ldots, v_{K}$ correspond to the operations executed in branches $1, \ldots, K$. This can be expressed more succinctly as a weighted combination,
\begin{equation}
    \label{eq:cond_succ}
    \text{cond}(\pi, v_1,\ldots,v_K) = \sum_{i=1}^K \pi_i v_i = \langle \pi, v \rangle,
\end{equation}
where $\langle \cdot, \cdot \rangle$ denotes the inner product and $v = (v_{1}, \ldots, v_{K})^{T}$. This formulation directly generalizes the binary masking operation from \autoref{eq:maskif}.

\paragraph{Differentiable multi-branch selection} Without loss of generality, assume the predicate takes the form of an $\argmax$ operation,
\begin{equation}
    \label{eq:argmax}
    \pi(x) = \argmax_{p \in \{e_{1}, \dots, e_{K}\}}\langle p, x \rangle.
\end{equation}
As in the binary case, this predicate is non-smooth because it can only take discrete values from ${e_{1}, \ldots, e_{K}}$. To construct a differentiable surrogate, we replace the discrete $\argmax$ with the softmax function,
\begin{equation}
    \text{softmax}(x)_{i} =  \frac{\exp(x_{i}/\tau)}{\sum_{j=1}^K \exp(x_j/\tau)},
\end{equation}
where $\tau > 0$ is a temperature hyperparameter controlling the sharpness of the approximation.
The softmax function emerges naturally from \autoref{eq:argmax} through entropy regularization, where the temperature $\tau$ corresponds to the regularization weight. As $\tau \to 0^{+}$, the softmax converges to the discrete $\argmax$, while larger values of $\tau$ produce smoother, more uniform distributions over branches.
Importantly, the softmax maps inputs to the unit simplex $\Delta_{K-1}$, allowing interpretation as a probability distribution over execution branches. From this perspective, discrete branch selections correspond to Dirac measures concentrated on $e_{1}, \ldots, e_{K}$, while the softmax provides a continuous relaxation across the simplex.

\paragraph{Implementation and computational considerations} The differentiable $K$-branch control flow statement is implemented by substituting the softmax approximation,
\begin{equation}
\text{cond}(\text{softmax}(x), v_1, \ldots, v_{K}) = \langle \text{softmax}(x), v\rangle.
\label{eq:softmaxif}
\end{equation}
As with binary control flow, all branches must be executed during both forward and backward passes to compute complete gradients. This necessitates implementing \autoref{eq:softmaxif} directly through matrix operations rather than using \texttt{else-if} or \texttt{switch} statements, which would cause standard AD engines to execute only the selected branch.
The computational cost scales linearly with the number of branches, but combinatorial explosion can occur with deeply nested structures. Control flow statements with fewer branches are preferable, and heavy nesting should be avoided. This again aligns well with typical ABM design principles, where individual agents employ simple decision-making processes corresponding to shallow control flows with low branching factors.

Multi-branch control flow structures naturally connect to discrete optimization problems. When agents must select from a finite set of actions, this can be formulated as an $\argmax$ operation (Equation \eqref{eq:argmax}). The softmax function therefore provides a principled approach for constructing differentiable surrogates for discrete action selection, enabling gradient-based learning in multi-agent environments with discrete choice spaces.

\subsection{Differentiable Discrete Randomness}
\label{sec:discrete_randomness}

Beyond control flow structures, ABMs frequently employ discrete random sampling to model stochastic agent behaviours. Most discrete probability distributions in ABMs are categorical distributions, to which we restrict our focus. For clarity, we primarily discuss Bernoulli distributions, though all methods generalize to Poisson and categorical distributions with arbitrary finite support. Unless otherwise specified, we treat samples from categorical distributions as one-hot encodings.

\paragraph{The Incompatibility of Standard Reparameterisation}

The fundamental challenge of differentiating through discrete randomness becomes apparent when considering the standard reparameterisation trick applied to a Bernoulli random variable $X \sim \text{Bern}(p)$. The reparameterisation proceeds by sampling $U \sim \text{Uniform}(0, 1)$ and computing $X = \mathbbm{1}_{U \leq p}$, where $\mathbbm{1}$ denotes the indicator function.

Whilst this reparameterisation enables sampling, it fails catastrophically for gradient computation. The gradient of the indicator function is zero almost everywhere and discontinuous at $p$, despite the fact that $\frac{d}{dp}\mathbb{E}[X] = 1 \neq 0$. Consequently, gradients computed from individual samples are completely uninformative and cannot be aggregated via Monte Carlo estimation to recover the true gradient of the expectation.

The core insight behind surrogate gradient methods is to replace these non-differentiable discrete sampling procedures with differentiable surrogates during the backward pass, while preserving the original discrete samples during the forward pass.

\subsubsection{The Straight-Through estimator}
\label{sec:st_estimator}

The simplest surrogate gradient approach replaces discrete sampling operations with the identity function during gradient computation. This straight-through (ST) estimator, popularised by \cite{bengio_estimating_2013}, derives its name from the fact that gradients pass ``straight through'' the sampling operation as if it were transparent.

Formally, for a categorical random vector $X \sim \text{Cat}(\pi)$, the ST estimator can be expressed as
\begin{equation}
    X^{\dagger} \sim \text{Cat}(\pi) + \pi - \bar{\pi},
\end{equation}
where $\bar{\pi}$ represents a stopped gradient (detached from the computational graph). During the forward pass, $X^{\dagger} = X$ since the additive term equals zero. However, during the backward pass, gradients flow through $\pi$ as if $X^{\dagger} = \pi$.
This reveals the ST estimator's fundamental approximation: it treats the discrete sample as if it were the continuous expectation $\mathbb{E}[\text{Cat}(\pi)] = \pi$ during gradient computation. The estimator essentially replaces the categorical distribution with a Dirac distribution centred at its mean.

\paragraph{Bias analysis and applicability} The bias of the ST estimator depends critically on the relationship between $f(X)$ and $f(\mathbb{E}[X])$ for the downstream function $f$. When approximating $\frac{d}{d\pi}\mathbb{E}[f(X)]$, the ST estimator exhibits low bias when $\mathbb{E}[f(X)] \approx f(\mathbb{E}[X])$ (i.e., when $f$ is approximately linear) and high bias otherwise.
This characteristic makes the ST estimator particularly suitable for ABMs with homogeneous agents that update their states in ways similar to mean-field approximations. In such scenarios, individual agent randomness often averages out across the population, making the mean-field approximation inherent in the ST estimator reasonable.
However, for ABMs where discrete stochastic choices lead to significantly different behavioural trajectories or where heterogeneity is crucial, the ST estimator's bias can be substantial, necessitating more sophisticated surrogate gradient methods.

\subsubsection{The Gumbel-Softmax Estimator}

Whilst the ST estimator provides a simple approach, more principled methods exist that better preserve the stochastic nature of discrete sampling. The Gumbel-Softmax (GS) estimator builds upon alternative reparameterisations of categorical distributions.
Recall that one may sample from $X \sim \text{Cat}(\pi)$ using the Gumbel-Max trick,
\begin{equation}
    X = \argmax_{p \in \{e_{1}, \dots e_{K}\}} \log\left(\sum_{i=1}^{K}\pi_{i}e_{i}\right) + \sum^{K}_{i=1}G_{i}e_{i}, \text{ where } G_{i} \sim \text{Gumbel}(0, 1),
\end{equation}
which reparameterises categorical sampling using Gumbel random variables and the $\argmax$ operator. As established in \autoref{sec:if_statements}, $\argmax$ operations over discrete sets are non-smooth. However, following the same approach as for control flow structures, we may replace the $\argmax$ function with the softmax function to obtain a smooth approximation for use during the tangent pass:
\begin{equation}
    \label{eq:gsoftmax}
    X \approx \text{softmax}\left(\frac{\log\pi + G}{\tau}\right).
\end{equation}
\autoref{eq:gsoftmax} defines the Gumbel-softmax estimator, first introduced by \cite{maddison_concrete_2016, jang_categorical_2016}. The surrogate gradient method that replaces categorical sampling procedures with the GS estimator during the tangent pass is known as the straight-through GS estimator in the AD literature.

\paragraph{Temperature and the bias-variance trade-off} The temperature parameter $\tau > 0$ controls the degree of smoothing applied to the categorical distribution. As $\tau \to 0^{+}$, the GS estimator converges to the exact Gumbel-Max trick and therefore to true categorical sampling. However, this temperature parameter introduces a fundamental bias-variance trade-off:
\begin{enumerate}
    \item Low temperature ($\tau \to 0^{+}$): The GS estimator closely resembles sampling from $\text{Cat}(\pi)$, resulting in low bias. However, the Gumbel noise passed to the softmax function is amplified, leading to high variance in gradient estimates.
    \item High temperature ($\tau \gg 0$): The impact of Gumbel noise is reduced, decreasing variance in gradient estimates. However, this increases bias by over-smoothing the categorical distribution.
\end{enumerate}
This trade-off may require tuning $\tau$ for each specific application.

\paragraph{Empirical illustration: Bernoulli random walk}

\begin{figure}
    \centering
    \includegraphics[width=0.75\linewidth]{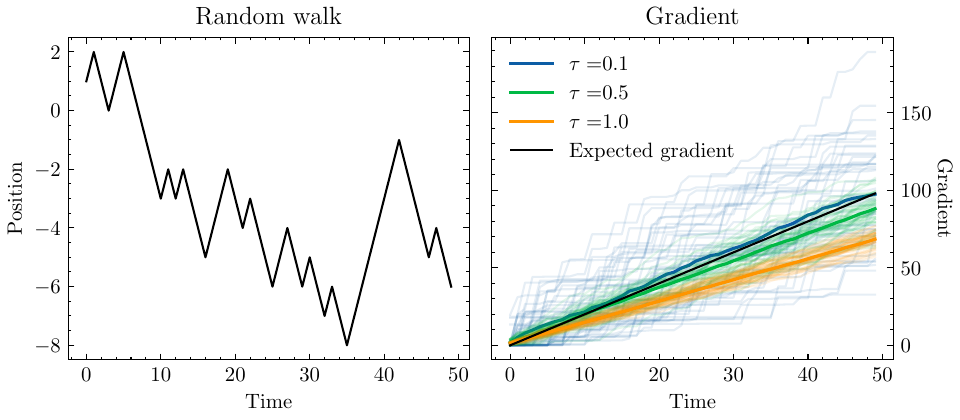}
    \caption{The left panel visualises a single simulation of the Bernoulli random walk (\autoref{eq:walk}) for $p=0.4$. The right panel depicts GS estimates of $d\mathbb{E}[X_{t}]/dp$ for $t=0, \ldots, 50$ under different temperature settings. Faint lines correspond to GS estimates from single simulation runs, whilst solid coloured lines show empirical averages across multiple runs.}
    \label{fig:gs_random_walk}
\end{figure}

To illustrate the effects of temperature, consider the Bernoulli random walk:
\begin{equation}
    X_{t} = X_{t-1} + (2\cdot \text{Bern}(p) - 1).
    \label{eq:walk}
\end{equation}
Figure \ref{fig:gs_random_walk} demonstrates the bias-variance trade-off empirically. The left panel shows a single simulation of the random walk for $p=0.4$, whilst the right panel compares GS estimates of $\frac{\mathrm{d}\mathbb{E}[X_{t}]}{dp}$ under different temperature settings against the analytical gradient $\frac{\mathrm{d}\mathbb{E}[X_{t}]}{\mathrm{d}p} = 2t$.
For $\tau = 0.1$, the empirical average of GS estimates closely approximates the exact gradient for all $t$, demonstrating low bias. However, individual estimates exhibit extreme variability, indicating high variance. Conversely, with $\tau = 1.0$, individual estimates are well-clustered (low variance), but the empirical average progressively diverges from the true gradient as $t$ increases, showing how bias compounds over time. The intermediate setting $\tau = 0.5$ achieves a desirable balance, exhibiting both low variance and low bias across all time steps.

\paragraph{Limitations of surrogate gradient methods}
Both the ST and GS estimators represent a particular methodological approach to handling discrete randomness: they modify the backward pass whilst preserving the forward pass, accepting some degree of bias in exchange for gradient flow. 

An alternative approach asks whether we can compute unbiased gradients of expectations involving discrete randomness without resorting to biased surrogates. This leads us to consider pathwise methods that directly tackle the mathematical challenge of differentiating through discrete jumps.

\subsubsection{Stochastic Derivatives: Extending Pathwise Gradients to Discrete Randomness}

Smoothed perturbation analysis (SPA) \citep{gong_smoothed_1987,ho_perturbation_1991} offers a fundamentally different approach that provides unbiased pathwise-type gradients for expectations involving discrete random choices. Unlike surrogate gradient methods that approximate discrete operations with smooth surrogates, SPA directly addresses the mathematical structure of discontinuous functions to compute exact derivatives.

The approach has been integrated into an AD framework by \cite{arya_automatic_2022} through the \textsc{Julia} programming language via the \texttt{StochasticAD.jl}\footnote{https://github.com/gaurav-arya/StochasticAD.jl} software package, enabling unbiased differentiation of programs that mix continuous and discrete randomness. For an in-depth discussion of the properties of this estimator we refer to \cite{arya_automatic_2022}.

\paragraph{Mathematical foundation} Consider a stochastic program $X(p)$ that may involve discrete random choices. Our objective is to evaluate
\begin{equation}
\frac{\mathrm{d}}{\mathrm{d}p}\,\mathbb{E}[X(p)]
  = \lim_{\varepsilon \to 0}
      \frac{\mathbb{E}\left[X(p+\varepsilon)-X(p)\right]}{\varepsilon}.
\end{equation}
For programs that vary smoothly in $p$, one simply employs the pointwise derivative $\partial X/\partial p$. However, discrete randomness introduces fundamental complications: when nudging the parameter from $p$ to $p+\varepsilon$, the function $X$ remains unchanged on almost every sample path, yielding $X(p+\varepsilon)-X(p)=0$ with probability one. Yet on the $\mathcal{O}(\varepsilon)$ fraction of paths where a discrete jump occurs, the change is of order $\mathcal{O}(1)$.

SPA captures both the infinitesimal smooth changes and the discrete jump contributions through the \textit{stochastic triple} object,
\begin{equation}
    (\delta,\;w,\;Y), \qquad
    \delta\in\mathbb{R},\;w\ge0,\;Y\in\mathrm{Range}(X),
\end{equation}
which obeys
\begin{equation}\label{eq:spa-estimator}
\frac{\mathrm{d}}{\mathrm{d}p}\,\mathbb{E}[X(p)]
    = \mathbb{E}\left[\delta + w\,\left(Y-X(p)\right)\right].
\end{equation}
The triple components have intuitive interpretations:
\begin{enumerate}
    \item $\delta$ is the usual infinitesimal contribution (zero for a purely discrete node);
    \item $w$ is an infinitesimal probability mass, of order $O(\varepsilon)$, that a discrete jump occurs; and
    \item $Y$ is the alternative value attained when the jump occurs.
\end{enumerate}

\paragraph{Example: a Bernoulli coin} 
\label{sec:bernoulli_example}

To illustrate how SPA converts a rare but finite jump into a usable derivative, consider a coin-flip simulator with parameter $p$. Sampling proceeds by drawing $\xi \sim U([0,1])$ and returning a Boolean outcome $x$ which equals 1 if $\xi$ is less than $p$, and 0 otherwise.

When we infinitesimally nudge the parameter $p$ whilst keeping $\xi$ fixed, most paths behave smoothly. For $\xi < p$ (heads) and increasing $p$ by $\varepsilon$, nothing changes as the sample remains heads. Similarly, for $\xi > p$ (tails) and decreasing $p$ by $\varepsilon$, the outcome remains unchanged.

The crucial behaviour occurs in a thin slab of width $\varepsilon$ adjacent to the threshold at $p$. For a right perturbation $p \mapsto p+\varepsilon$ with $\varepsilon > 0$, a ``tails'' path ($x=0$) flips to ``heads'' if and only if $\xi$ falls in the interval $(p, p+\varepsilon)$. This occurs with conditional probability $\varepsilon/(1-p)$, leading SPA to assign weight $w^+ = \frac{1}{1-p},\mathbbm{1}_{[x=0]}$ and alternative value $Y^+ = 1$.

For a left perturbation $p \mapsto p-\varepsilon$, a ``heads'' path ($x=1$) flips to ``tails'' when $\xi \in (p-\varepsilon, p)$, occurring with conditional probability $\varepsilon/p$. Here, SPA assigns weight $w^- = \frac{1}{p},\mathbf{1}_{[x=1]}$ and alternative value $Y^- = 0$.

In both cases the classical infinitesimal part is zero ($\delta = 0$) because
the output is piecewise constant.  Plugging the triples
$(\delta,w^\pm,Y^\pm)$ into the SPA estimator~\eqref{eq:spa-estimator}
gives, e.g.\ for the right-hand side,
\begin{equation}
\mathbb{E}\left[w^+\,(Y^+ - x)\right] \;=\; \frac{1}{1-p}\,\mathbb{P}[x=0] \;=\; \frac{1}{1-p}\,(1-p) \;=\; 1,
\end{equation}
with an analogous calculation for the left perturbation yielding the same value.
Hence SPA correctly recovers
$\tfrac{\mathrm{d}}{\mathrm{d}p}\,\mathbb{E}[x] = 1$,
matching the analytic derivative of $\mathbb{E}[x]=p$.

We refer to \cite{arya_automatic_2022} for a complete derivation of rules for other commonly used discrete probability distributions.

\paragraph{Composition (chain rule) for stochastic triples.}

With a stochastic–derivative triple $(\delta,w,Y)$ attached to every operation, the derivative of an arbitrary program is obtained by repeatedly composing those triples, mirroring the core idea of classical automatic differentiation. All we need is a rule that merges triples across a function call boundary.

Consider a program that evaluates an inner subroutine $a = f(p)$ followed by an outer subroutine $X = g(a)$, with respective triples $(\delta_f, w_f, Y_f)$ and $(\delta_g, w_g, Y_g)$. Since SPA perturbs all random draws by the same infinitesimal $\epsilon$, the probability that two nodes experience a jump is of order $\mathcal{O}(\epsilon^{2})$. Terms of such order make a negligible contribution to the stochastic derivative and so we may focus on cases where only a single node undergoes a discrete jump and deploy the following chain rule:

\begin{itemize}[leftmargin=1.5em,itemsep=.4em]
\item
\textbf{Jump in the outer node.}  
The inner call is smooth but the outer call flips to $Y_g$,
\begin{equation}
\label{eq:comp_outer}
\left(
  \delta_f\,\delta_g + w_f\,[\,g(Y_f) - g(a)\,],\; w_g,\; Y_g
\right).
\end{equation}

\item
\textbf{Jump in the inner node.}  
The inner call supplies the alternative output $Y_f$; the outer call responds
infinitesimally around that new input,
\begin{equation}\label{eq:comp_inner}
\left(
  \delta_f\,\delta_g,\; w_f,\; g(Y_f)
\right).
\end{equation}
\end{itemize}
These composition rules reduce to the classical chain rule when $w_f = w_g = 0$.

\paragraph{Computational scalability: pruning and smoothing}

A program with $k$ discrete nodes has $2^{k}-1$ possible jump combinations after infinitesimal perturbation, making enumeration computationally intractable for large $k$. In ABMs, sampling the state of $N$ agents at one time step already sets $k = N$, potentially rendering the method infeasible.

\paragraph{Pruning strategy}

\texttt{StochasticAD.jl} sidesteps this computational wall with a \emph{pruning} scheme that limits the search to a single alternative per \emph{tag}.  Every fresh random draw is stamped with a tag, and --- thanks to Julia’s multiple dispatch --- this tag is automatically propagated to all downstream operations that depend on that draw. During gradient evaluation, each tag $t$ contributes candidate jumps $\{(\delta,w,Y)\}_{t}$, from which one is retained randomly with probability proportional to its weight. Because the retention probability is proportional to the local weight, the expectation over pruning outcomes produces an unbiased estimate of the gradient.

\paragraph{Smoothing}

Pruning leaves the estimator unbiased but can incur substantial variance. A useful variance–reduction alternative is \emph{smoothing}, which replaces the random jump at each node by its conditional mean, effectively applying a
mean-field correction.

Formally, if a node has stochastic-derivative triple
$(\delta,w,Y)$, we define its \emph{smoothed} derivative as \citep{arya_automatic_2022}
\begin{equation}\label{eq:smoothing}
\tilde{\delta}
  \;=\;
  \mathbb{E}\left[\,
      \delta \;+\; w\left(Y - X(p)\right)
      \,\bigm|\,X(p)\right],
\end{equation}
The finite jump $Y-X(p)$ is averaged into an $\mathcal{O}(\varepsilon)$ infinitesimal, so the downstream graph once again obeys the ordinary chain rule. Consequently, any AD backend can run in standard forward or reverse-mode, with lower variance albeit at the cost of a bias that vanishes in locally linear regions.

When applying smoothing to our Bernoulli example, the right-hand triple
\((0,\;1/(1-p)\,\mathbbm 1_{x=0},\;1)\) becomes

\begin{equation}
\tilde{\delta}
  \;=\;
  \mathbb{E}\left[w\,(1-x)\mid x\right]
  \;=\;
  \frac{1}{1-p}\,\mathbbm 1_{x=0}.
\end{equation}
An analogous expression may be derived for the left-hand triple. By combining both sides we recover the ST gradient estimator.

\section{Differentiable Agent-Based Models}
\label{sec:diff_abms}

We now return to the agent-based models described in \autoref{sec:abms}. In what follows, we construct differentiable implementations of three representative ABMs using the techniques introduced in \autoref{sec:autodiff}: Axtell's model of firms (AMOF), Sugarscape, and the SIR epidemiological model. Each model presents distinct challenges for differentiation, ranging from discrete optimization (AMOF) to spatial movement and resource allocation (Sugarscape) to stochastic state transitions (SIR).

Before proceeding, we stress that both surrogate gradient methods and stochastic derivatives leave the primal pass unchanged. As a result, the original discrete dynamics and stochastic elements that characterise each model's behaviour remain intact. In fact, smooth approximations are only introduced during the tangent pass with the aim of enabling the gradient-based calibration procedures that will be presented in \autoref{sec:calib}.

\subsection{Differentiable Axtell Model of Firms}

We begin with AMOF, where the primary non-differentiable element is the discrete firm selection process in which agents make choices between staying, switching, or founding firms based on utility maximization.

The key challenge in making AMOF differentiable lies in the discrete decision process at the end of each time step wherein agents choose a firm to join. Recall that each agent $i$ computes utilities $U^{\star}_i(f)$ for all firms $f \in \mathcal{F}_i$, where $\mathcal{F}_{i}$ contains the agent $i$'s current firm, the firms employing agent $i$'s friends, and the new firm that could be founded by agent $i$ (if it has not found a new firm already). Agent $i$ then selects the firm that maximises its utility via an $\argmax$ operation,
\begin{equation}
    \label{eq:firm_selection}
    f_{i}^{\star} = \arg\max_{f \in \mathcal{F}_{i}} \: U_{i}^{\star}(f).
\end{equation}
This discrete optimization over a finite set introduces non-differentiability, preventing gradient flow through the firm selection process to the underlying model parameters (Table \ref{tab:properties_axtell}).

\paragraph{Surrogate gradient implementation}
Following the multi-branch control flow approach from \autoref{sec:multi_branch}, we treat each firm $f \in \mathcal{F}{i}$ as a component of a $|\mathcal{F}{i}|$-dimensional one-hot vector and replace the discrete $\argmax$ with a softmax approximation during the tangent pass.
Specifically, we replace the chosen firm $f_{i}^{\star}$ in the tangent pass with an exponentially weighted mixture of firms,
\begin{equation}
    \label{eq:mixed_firm}
    \tilde{f}_{i} = \frac{\exp(U^{\star}_i)}{\sum_{f \in \mathcal{F}_i} \exp(U^{\star}_i(f))},
\end{equation}
where $U^{\star}_{i} = (U_{i}(f))_{f \in \mathcal{F}_{i}}$ is the vector of utilities associated for each potential firm $f \in \mathcal{F}_{i}$ and $\tau>0$ is a temperature parameter controlling the sharpness of the approximation.

This formulation preserves the discrete dynamics during the forward pass—each agent still makes a definitive choice according to \autoref{eq:firm_selection}—while enabling gradient computation during the backward pass through the continuous softmax approximation.

The softmax approximation necessitates corresponding updates to all model components that depend on firm membership:

\paragraph{Firm size updates} During the forward pass, firm size $|f|$ is incremented by one when agent $i$ joins firm $f$. In the tangent pass, this discrete update is replaced by:
\begin{equation}
    \label{eq:firm_size}
    |f| \leftarrow |f| + \langle f , \tilde{f}_{i} \rangle = |f| + \frac{\exp (U^{\star}_{i}(f))}{\sum_{f^{\prime} \in \mathcal{F}_{i}}\exp(U^{\star}_{i}(f^{\prime}))},
\end{equation}
where $\tilde{f}_{i,f}$ represents the probability that agent $i$ chooses firm $f$ under the softmax distribution.

\paragraph{Effort level updates} Similarly, agent $i$'s effort level for the next time step is set to $e^{\star}_{i}(f_{i}^{\star})$ during the forward pass, but becomes an exponentially weighted average during the tangent pass,
\begin{equation}
    \label{eq:mixed_effort}
    e_{i}(t+1) = \langle e_{i}^{\star}, \tilde{f} \rangle = \frac{\sum_{f \in \mathcal{F}_{i}}\exp(U_{i}^{\star}(f))e^{\star}_{i}(f)}{\sum_{f^{\prime}\in \mathcal{F}_{i}}\exp(U^{\star}_{i}(f^{\prime}))},
\end{equation}
where $e^{\star}_{i}(f)$ is the optimal effort level agent $i$ would choose if joining firm $f$.

\paragraph{Computational considerations} Note that the differences between the primal and tangent passes described by Equations \eqref{eq:mixed_firm}, \eqref{eq:firm_size}, and \eqref{eq:mixed_effort}, arise simply from one-hot encoding firms and replacing the $\argmax$ operation in \autoref{eq:firm_selection} with a softmax operation during the tangent pass via the surrogate gradient method. The introduction of these features comes at a cost in terms of both memory and time complexity. The memory required to simulate agent $i$'s behaviour on each time step using one-hot encoded firms scales linearly with the number of firms agent $i$ can join, $|\mathcal{F}_{i}|$. Likewise, the time complexity of the softmax operations detailed by Equations \eqref{eq:mixed_firm}, \eqref{eq:firm_size} and \eqref{eq:mixed_effort} scale linearly with $|\mathcal{F}_{i}|$. These computational burdens accumulate additively as the agent population increases. As a result, it is critical that the friendship network between agents is sparse, and $|\mathcal{F}_{i}|$ is small for each agent $i$.

\subsection{Differentiable Sugarscape Model}

We now demonstrate how to construct a differentiable version of Sugarscape using the surrogate gradient techniques from \autoref{sec:autodiff}. The primary non-differentiable elements in Sugarscape are: (1) the discrete cell selection process where agents choose movement destinations via $\argmax$ over visible cells, (2) the categorical sampling of vision ranges, and (3) the binary survival decisions based on sugar holdings.

\paragraph{Matrix-based representation of agent vision}

The key insight for making Sugarscape differentiable lies in reformulating the spatial movement decisions using matrix operations that can accommodate surrogate gradients. Recall that each agent searches through available cells in its vision range and moves to the cell with the highest sugar reserve—an optimization over a finite set whose contents depend on the agent's vision range, position, and other agents' positions.

To handle this systematically, we assume that agent vision ranges are bounded above by some fixed $V \in \mathbb{N}$. This allows us to represent any agent's neighborhood as a $(2V + 1) \times (2V + 1)$ matrix, where larger vision ranges are subsets of this maximum representation.

For an agent with vision radius $v \leq V$, we define a vision matrix $\mathbf{M}_v$ as
\begin{equation}
    (\mathbf M_v)_{ij} = 
    \begin{cases}
        1 & \text{if } |i - (V+1)| + |j - (V+1)| < v, \\
        0 & \text{otherwise},
    \end{cases}
    \quad \text{for } 1 \leq i,j \leq 2V+1.
\end{equation}
The central index $(V+1, V+1)$ corresponds to the cell $(x, y)$ currently occupied by the agent. If $(\mathbf{M}_{v})_{ij} = 1$, then cell $(x + (V + 1) - i, y + (V+1) - j)$ is currently visible to the agent. Similarly, if $(\mathbf{M}_{v})_{ij} = 0$, then cell  $(x + (V + 1) - i, y + (V+1) - j)$ is not visible to the agent. Note that every cell visible to the agent has a corresponding index in the matrix $\mathbf{M}_{v}$. Any cell in the environment without a corresponding index in the matrix $\mathbf{M}_{v}$ is not visible to the agent, since the cell lies beyond the maximum vision radius $V$ relative to the agent's current position. The vision matrices $\mathbf{M}_{1}, \mathbf{M}_{2}, \mathbf{M}_{3}$ for $V=3$ are depicted in Figure \ref{fig:vision_matrices}.

\begin{figure}
    \centering
    \scriptsize
    \definecolor{color1}{RGB}{0,0,255} 
\definecolor{color0}{RGB}{150,150,150} 
\begin{align*}
& \hspace{1.90cm} \mathbf M_{1} && \hspace{1.90cm} \mathbf M_{2} && \hspace{1.90cm} \mathbf M_{3} \\
&\begin{pmatrix}
\textcolor{color0}{0} & \textcolor{color0}{0} & \textcolor{color0}{0} & \textcolor{color0}{0} & \textcolor{color0}{0} & \textcolor{color0}{0} & \textcolor{color0}{0} \\
\textcolor{color0}{0} & \textcolor{color0}{0} & \textcolor{color0}{0} & \textcolor{color0}{0} & \textcolor{color0}{0} & \textcolor{color0}{0} & \textcolor{color0}{0} \\
\textcolor{color0}{0} & \textcolor{color0}{0} & \textcolor{color0}{0} & \textcolor{color1}{1} & \textcolor{color0}{0} & \textcolor{color0}{0} & \textcolor{color0}{0} \\
\textcolor{color0}{0} & \textcolor{color0}{0} & \textcolor{color1}{1} & \textcolor{color1}{1} & \textcolor{color1}{1} & \textcolor{color0}{0} & \textcolor{color0}{0} \\
\textcolor{color0}{0} & \textcolor{color0}{0} & \textcolor{color0}{0} & \textcolor{color1}{1} & \textcolor{color0}{0} & \textcolor{color0}{0} & \textcolor{color0}{0} \\
\textcolor{color0}{0} & \textcolor{color0}{0} & \textcolor{color0}{0} & \textcolor{color0}{0} & \textcolor{color0}{0} & \textcolor{color0}{0} & \textcolor{color0}{0} \\
\textcolor{color0}{0} & \textcolor{color0}{0} & \textcolor{color0}{0} & \textcolor{color0}{0} & \textcolor{color0}{0} & \textcolor{color0}{0} & \textcolor{color0}{0}
\end{pmatrix}, &&
\begin{pmatrix}
\textcolor{color0}{0} & \textcolor{color0}{0} & \textcolor{color0}{0} & \textcolor{color0}{0} & \textcolor{color0}{0} & \textcolor{color0}{0} & \textcolor{color0}{0} \\
\textcolor{color0}{0} & \textcolor{color0}{0} & \textcolor{color0}{0} & \textcolor{color1}{1} & \textcolor{color0}{0} & \textcolor{color0}{0} & \textcolor{color0}{0} \\
\textcolor{color0}{0} & \textcolor{color0}{0} & \textcolor{color1}{1} & \textcolor{color1}{1} & \textcolor{color1}{1} & \textcolor{color0}{0} & \textcolor{color0}{0} \\
\textcolor{color0}{0} & \textcolor{color1}{1} & \textcolor{color1}{1} & \textcolor{color1}{1} & \textcolor{color1}{1} & \textcolor{color1}{1} & \textcolor{color0}{0} \\
\textcolor{color0}{0} & \textcolor{color0}{0} & \textcolor{color1}{1} & \textcolor{color1}{1} & \textcolor{color1}{1} & \textcolor{color0}{0} & \textcolor{color0}{0} \\
\textcolor{color0}{0} & \textcolor{color0}{0} & \textcolor{color0}{0} & \textcolor{color1}{1} & \textcolor{color0}{0} & \textcolor{color0}{0} & \textcolor{color0}{0} \\
\textcolor{color0}{0} & \textcolor{color0}{0} & \textcolor{color0}{0} & \textcolor{color0}{0} & \textcolor{color0}{0} & \textcolor{color0}{0} & \textcolor{color0}{0}
\end{pmatrix}, &&
\begin{pmatrix}
\textcolor{color0}{0} & \textcolor{color0}{0} & \textcolor{color0}{0} & \textcolor{color1}{1} & \textcolor{color0}{0} & \textcolor{color0}{0} & \textcolor{color0}{0} \\
\textcolor{color0}{0} & \textcolor{color0}{0} & \textcolor{color1}{1} & \textcolor{color1}{1} & \textcolor{color1}{1} & \textcolor{color0}{0} & \textcolor{color0}{0} \\
\textcolor{color0}{0} & \textcolor{color1}{1} & \textcolor{color1}{1} & \textcolor{color1}{1} & \textcolor{color1}{1} & \textcolor{color1}{1} & \textcolor{color0}{0} \\
\textcolor{color1}{1} & \textcolor{color1}{1} & \textcolor{color1}{1} & \textcolor{color1}{1} & \textcolor{color1}{1} & \textcolor{color1}{1} & \textcolor{color1}{1} \\
\textcolor{color0}{0} & \textcolor{color1}{1} & \textcolor{color1}{1} & \textcolor{color1}{1} & \textcolor{color1}{1} & \textcolor{color1}{1} & \textcolor{color0}{0} \\
\textcolor{color0}{0} & \textcolor{color0}{0} & \textcolor{color1}{1} & \textcolor{color1}{1} & \textcolor{color1}{1} & \textcolor{color0}{0} & \textcolor{color0}{0} \\
\textcolor{color0}{0} & \textcolor{color0}{0} & \textcolor{color0}{0} & \textcolor{color1}{1} & \textcolor{color0}{0} & \textcolor{color0}{0} & \textcolor{color0}{0}
\end{pmatrix}.
\end{align*}
    \caption{The vision matrices $\mathbf{M}_{1}, \mathbf{M}_{2}, \mathbf{M}_{3}$ for $V=3$.}
    \label{fig:vision_matrices}
\end{figure}

\begin{figure}[ht]
    \centering
    \includegraphics[width=0.7\linewidth]{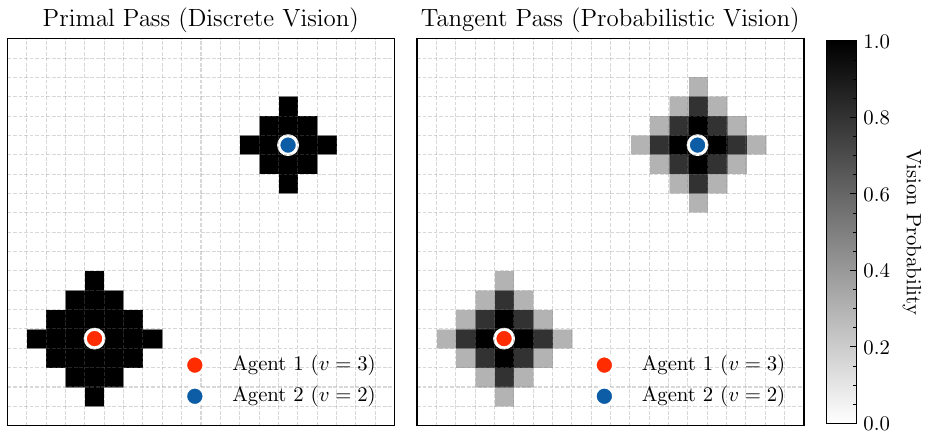}
    \caption{The vision ranges used by two agents during the primal (left) and tangent (right) passes. In the primal pass, a single discrete vision matrix is used, while in the tangent pass, an average over all possible vision matrices ($\mathcal{V} = \{1, \dots, 4\}$) is used.}
    \label{fig:sugarscape_vision}
\end{figure}

\paragraph{Handling categorical vision sampling}
As described, in Section \ref{sec:sugarscape}, the vision range of each agent is sampled from a categorical distribution $\text{Cat}(p)$ over a finite set of ranges $\mathcal{V} = \{v_{1}, \cdots, v_{k}\}$ each with corresponding vision matrix $\mathbf{M}_{v_{1}}, \cdots, \mathbf{M}_{v_{k}}$. During the primal pass, agent $i$ adopts the vision matrix $\mathbf{M}_{v_{i}}$ associated with its vision range $v_{i}$. To accommodate categorical sampling during the tangent pass, we use the ST estimator. As a result, the vision matrix $\mathbf{\tilde{M}}$ adopted by agent $i$ during the tangent pass is a mixture of vision matrices weighted by $p$:
\begin{equation}
\tilde {\mathbf{M}} = \sum_{v \in \mathcal{V}} p_v \mathbf{M}_v.
\end{equation}
In other words, each agent uses the average vision matrix, as defined by $p$, during the tangent pass. Figure \ref{fig:sugarscape_vision} visualises the vision ranges used by two different agents during the primal and tangent passes.

\paragraph{Differentiable cell selection} With vision matrices established, we can reformulate cell selection as matrix operations. For agent $i$ at time $t$, we construct local matrices describing the agent's immediate environment,
\begin{enumerate}
    \item \textit{Local sugar matrix} $\mathbf{S}_{i, t}$: sugar reserves in the agent's vicinity,
    \item \textit{Local occupancy matrix} $\mathbf{O}_{i, t}$: occupancy status of nearby cells.
\end{enumerate}
These are defined by indexing into the global environment,
\begin{equation}
    \begin{aligned}
        (\mathbf{S}_{i, t})_{k, l} &= s(x_{i}(t) + (V+1) - k, \; y_{i}(t) + (V+1) - l,\; t),\\
        (\mathbf{O}_{i, t})_{k, l} &= 1 - o(x_{i}(t) + (V+1) - k, \; y_{i}(t) + (V+1) - l,\; t),
    \end{aligned}
\end{equation}
where $(x_{i}(t), y_{i}(t))$ is agent $i$'s current position, $s(x,y,t)$ denotes the sugar available on cell $(x,y)$ at time $t$, and $o(x,y,t)$ is the binary variable denoting whether cell $(x,y)$ is occupied at time $t$.

Agent $i$ then computes a score matrix by combining vision, sugar, and occupancy information,
\begin{equation}
    \label{eq:score}
    \mathbf{Z}_{i, t} = \mathbf{S}_{i, t} \odot \mathbf{O}_{i, t} \odot \mathbf{M}_{v_{i}},
\end{equation}
where $\odot$ denotes element-wise multiplication. This masks out cells that are either invisible or occupied, leaving only viable movement options with their corresponding sugar values.
The optimal cell selection becomes
\begin{equation}
    \label{eq:best_cell}
    \argmax_{k, l} \: (\mathbf{Z}_{i, t})_{kl}.
\end{equation}

\paragraph{Surrogate gradient implementation} 
Following our multi-branch approach from Section \ref{sec:multi_branch}, we represent the $\argmax$ solution as a one-hot matrix $\mathbf{Z}^{\star}_{i, t}$,
\begin{equation}
    (\mathbf{Z}^{\star}_{i,t})_{ab} =
    \begin{cases}
        1 \quad \text{if } (a, b) = \argmax_{k, l} \: (\mathbf{Z}_{i, t})_{kl} \\
        0 \quad \text{otherwise.}
    \end{cases}
\end{equation}
During the tangent pass, we replace this discrete selection with a softmax approximation
\begin{equation}
    (\mathbf{\tilde{Z}}_{i,t})_{lk} = \frac{\exp((\mathbf{Z}_{i,t})_{lk})}{\sum_{a,b}\exp((\mathbf{Z}_{i,t})_{ab})}.
\end{equation}
This transforms the discrete cell choice into a probability distribution over all possible destinations for the tangent pass.

\paragraph{Position and wealth updates}

\begin{figure}
    \centering
    \scriptsize
    \definecolor{color1}{RGB}{0,0,255} 
\definecolor{color0}{RGB}{150,150,150} 
\begin{align*}
& \hspace{2.6cm} \Delta_{x} && \hspace{2.4cm}  \Delta_{y} \\
&\begin{pmatrix}
\textcolor{color0}{-3} & \textcolor{color0}{-3} & \textcolor{color0}{-3} & \textcolor{color0}{-3} & \textcolor{color0}{-3} & \textcolor{color0}{-3} & \textcolor{color0}{-3} \\
\textcolor{color0}{-2} & \textcolor{color0}{-2} & \textcolor{color0}{-2} & \textcolor{color0}{-2} & \textcolor{color0}{-2} & \textcolor{color0}{-2} & \textcolor{color0}{-2} \\
\textcolor{color0}{-1} & \textcolor{color0}{-1} & \textcolor{color0}{-1} & \textcolor{color0}{-1} & \textcolor{color0}{-1} & \textcolor{color0}{-1} & \textcolor{color0}{-1} \\
\textcolor{color0}{0} & \textcolor{color0}{0} & \textcolor{color0}{0} & \textcolor{color0}{0} & \textcolor{color0}{0} & \textcolor{color0}{0} & \textcolor{color0}{0} \\
\textcolor{color0}{\mathord{+}1} & \textcolor{color0}{\mathord{+}1} & \textcolor{color0}{\mathord{+}1} & \textcolor{color0}{\mathord{+}1} & \textcolor{color0}{\mathord{+}1} & \textcolor{color0}{\mathord{+}1} & \textcolor{color0}{\mathord{+}1} \\
\textcolor{color0}{\mathord{+}2} & \textcolor{color0}{\mathord{+}2} & \textcolor{color0}{\mathord{+}2} & \textcolor{color0}{\mathord{+}2} & \textcolor{color0}{\mathord{+}2} & \textcolor{color0}{\mathord{+}2} & \textcolor{color0}{\mathord{+}2} \\
\textcolor{color0}{\mathord{+}3} & \textcolor{color0}{\mathord{+}3} & \textcolor{color0}{\mathord{+}3} & \textcolor{color0}{\mathord{+}3} & \textcolor{color0}{3} & \textcolor{color0}{\mathord{+}3} & \textcolor{color0}{\mathord{+}3}
\end{pmatrix}, &&
\begin{pmatrix}
\textcolor{color0}{-3} & \textcolor{color0}{-2} & \textcolor{color0}{-1} & \textcolor{color0}{0} & \textcolor{color0}{\mathord{+}1} & \textcolor{color0}{\mathord{+}2} & \textcolor{color0}{\mathord{+}3} \\
\textcolor{color0}{-3} & \textcolor{color0}{-2} & \textcolor{color0}{-1} & \textcolor{color0}{0} & \textcolor{color0}{\mathord{+}1} & \textcolor{color0}{\mathord{+}2} & \textcolor{color0}{\mathord{+}3}  \\
\textcolor{color0}{-3} & \textcolor{color0}{-2} & \textcolor{color0}{-1} & \textcolor{color0}{0} & \textcolor{color0}{\mathord{+}1} & \textcolor{color0}{\mathord{+}2} & \textcolor{color0}{\mathord{+}3} \\
\textcolor{color0}{-3} & \textcolor{color0}{-2} & \textcolor{color0}{-1} & \textcolor{color0}{0} & \textcolor{color0}{\mathord{+}1} & \textcolor{color0}{\mathord{+}2} & \textcolor{color0}{\mathord{+}3}  \\
\textcolor{color0}{-3} & \textcolor{color0}{-2} & \textcolor{color0}{-1} & \textcolor{color0}{0} & \textcolor{color0}{\mathord{+}1} & \textcolor{color0}{\mathord{+}2} & \textcolor{color0}{\mathord{+}3}  \\
\textcolor{color0}{-3} & \textcolor{color0}{-2} & \textcolor{color0}{-1} & \textcolor{color0}{0} & \textcolor{color0}{\mathord{+}1} & \textcolor{color0}{\mathord{+}2} & \textcolor{color0}{\mathord{+}3}  \\
\textcolor{color0}{-3} & \textcolor{color0}{-2} & \textcolor{color0}{-1} & \textcolor{color0}{0} & \textcolor{color0}{\mathord{+}1} & \textcolor{color0}{\mathord{+}2} & \textcolor{color0}{\mathord{+}3} 
\end{pmatrix}.
\end{align*}
    \caption{The shift matrices $\Delta_{x}$ and $\Delta_{y}$ for $V=3$.}
    \label{fig:shifts}
\end{figure}

To translate the matrix-based selection back to spatial coordinates, we use shift matrices $\Delta_{x}$ and $\Delta_{y}$ that encode the relative movements
\begin{equation}
    \begin{aligned}
        (\Delta_{x})_{ij} &= i - (V+1) \\
        (\Delta_{y})_{ij} &= j - (V+1).
    \end{aligned}
\end{equation}
Both shift matrices are displayed in Figure \ref{fig:shifts} for $V=3$. Agent positions update via Frobenius inner products,
\begin{equation}
    \begin{aligned}
        x_{i}(t+1) &= x_{i}(t) + \langle \mathbf{Z}^{\star}_{i,t}, \Delta_{x}\rangle \\
        y_{i}(t+1) &= y_{i}(t) + \langle \mathbf{Z}^{\star}_{i,t}, \Delta_{y}\rangle,
    \end{aligned}
\end{equation}

Similarly, sugar holdings update by harvesting from the selected cell,
\begin{equation}
    h_{i}(t+1) = h_{i}(t) + [\langle \mathbf{Z}^{\star}_{i, t}, \mathbf{S}_{i,t}\rangle - m_{i}] \cdot a_{i}(t),
\end{equation}
where $a_{i}(t)$ is a binary variable denoting whether agent $i$ is alive on time step $t$.

\paragraph{Handling agent survival} Agent survival introduces another non-differentiable element: the binary life-or-death decision based on whether $h_{i}(t+1) \leq 0$. This discrete predicate depends on model parameters through the sugar dynamics. We handle this using the control flow techniques from \autoref{sec:if_statements}, replacing the binary survival indicator with a sigmoid approximation during the tangent pass.

Both $\mathbf{S}_{i,t}$ and $\mathbf{O}_{i,t}$ are then modified to reflect agent $i$'s recent actions and are used to update the global matrices $\mathbf{S}_{t}$ and $\mathbf{O}_{t}$. Finally, after each agent has moved, the sugar reserves regenerate before the next time step begins. 

We emphasize that our implementation does not propagate second-order gradient terms arising from the interaction between agent movement and environment perception. For example, when computing gradients with respect to vision parameters, our method only accounts for their direct effect on what agents perceive at their chosen destinations. It ignores the indirect effect that altered vision parameters might change movement decisions, leading agents to different locations where they would observe different environmental states. These second-order terms are omitted because gradients cannot flow through the discrete indexing operations used to construct the local observation matrices $\mathbf{O}{i,t}$ and $\mathbf{S}{i,t}$. This simplification is analogous to practices in perturbation theory, where system responses are expanded around a reference trajectory and only the leading-order terms are retained. Our approach similarly captures the dominant sensitivity structure while neglecting higher-order corrections that would require significant computational overhead relative to their expected contribution.

\subsection{Differentiable SIR Model}
\label{sec:diff_sir}
As our last differentiable implementation, we now examine how to construct a differentiable version of the SIR epidemiological model introduced in \autoref{sec:sir_intro}. The SIR model presents several key differentiability challenges that exemplify the broader issues discussed in this paper: discrete state transitions governed by Bernoulli sampling, discrete policy timing controls, and stochastic agent behaviours. The basis of this implementation was first developed by \cite{chopra_differentiable_2023}.

\paragraph{Discrete state transition}
The core dynamics of the SIR model involve discrete state transitions that occur through Bernoulli sampling procedures. Consider a susceptible agent $i$ at time step $t-1$. The agent's infection status $I_i(t)$ at the next time step is determined by sampling from a Bernoulli distribution with success probability derived from the force of infection,
\begin{equation}
    \label{eq:infections}
    I_i(t) \sim \mathrm{Bern}(1 - \exp(-\lambda_i(t)\Delta t)).
\end{equation}
Similarly, for an infected agent $i$ at time step $t-1$, the recovery status $R_i(t)$ follows
\begin{equation}
    \label{eq:recovery}
    R_i(t) \sim \mathrm{Bern}(1 - \exp(-\gamma\Delta t)).
\end{equation}
Quarantine compliance introduces an additional layer of discrete sampling. During active quarantine periods, each agent's compliance decision $Q_i(t)$ is sampled as
\begin{equation}
    \label{eq:quarantine}
    Q_i(t) \sim \mathrm{Bern}(p_Q).
\end{equation}

Each of these discrete sampling operations breaks gradient flow and requires attention. The methods discussed in Section \ref{sec:discrete_randomness} can be applied to address these challenges, though their relative effectiveness may vary depending on the specific epidemiological context. We provide a comprehensive empirical comparison of these approaches in \autoref{sec:sir_gradients}.

\paragraph{Policy timing controls} Beyond discrete sampling, the SIR model incorporates policy interventions that activate and deactivate based on explicit timing parameters. Quarantine interventions are active only during the interval $[Q_{\text{start}}$, $Q_{\text{end}}]$,  while social distancing measures operate within $[D_{\text{start}}$, $D_{\text{end}}]$. These timing controls introduce discontinuous behaviour changes that correspond to gate functions of the form
\begin{equation}
    g(t) = \mathbbm{1}_{\{t_{\text{start}} \leq t \leq t_{\text{end}}\}}.
\end{equation}
Note that this gate function is simply the product of two step functions (one inverted): $g(t) = \mathbbm{1}_{{t \geq t_{\text{start}}}} \cdot \mathbbm{1}_{{t \leq t_{\text{end}}}}$.

\begin{figure}
    \centering
    \includegraphics[width=0.6\linewidth]{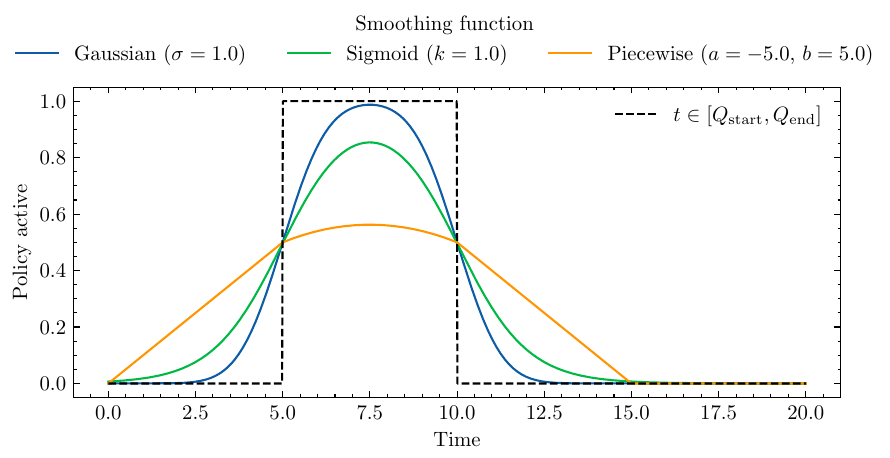}
    \caption{Smooth surrogates for the gate function corresponding to the predicate $Q_{\text{start}} \leq t \leq Q_{\text{end}}$ with $Q_{\text{start}} = 5$ and $Q_{\text{end}} = 10$. Three different smoothing approaches are shown: Gaussian CDF, sigmoid function, and piecewise linear approximations. Each provides a differentiable approximation to the discrete policy activation behaviour.}
    \label{fig:smoothing}
\end{figure}

Following the surrogate gradient approach outlined in Section \ref{sec:if_statements}, we replace these step functions with smooth approximations during gradient computation. Figure \ref{fig:smoothing} illustrates several smooth surrogates for a representative gate function. In our implementation, we adopt a surrogate constructed using the Gaussian cumulative distribution function with standard deviation $\sigma = 1$, which provides a good balance between smoothness and approximation accuracy. 

The combination of these techniques—smoothing discrete sampling operations and approximating gate functions—enables gradient-based optimization over all nine parameters of the SIR model, including both epidemiological parameters ($\beta$, $\gamma$, $I_0$) and policy intervention parameters ($Q_{\text{start}}$, $Q_{\text{end}}$, $p_Q$, $D_{\text{start}}$, $D_{\text{end}}$, $\alpha_D$).

\section{Empirical Validation of Differentiable Agent-Based Models}
\label{sec:grad_val}

Equipped with differentiable implementations for each of the ABMs introduced in \autoref{sec:abms}, we now empirically validate gradient estimates obtained through AD. Specifically, we evaluate the accuracy of AD gradients by comparing them against finite differences (FD), which serve as our ground truth baseline.

For each ABM, we seek to approximate 
\begin{equation}
    \nabla_{\theta}\mathbb{E}_{x \sim p(\cdot \mid \theta)}[\Phi(x)],
\end{equation}
where $x \in \mathcal{X}$ represents the model output, $p(\cdot \mid \theta)$ denotes the probability distribution over $\mathcal{X}$ induced by the model parameterized by $\theta$, and $\Phi$ maps each output $x$ to its emergent representation (following the notation established in \autoref{sec:abms}).

\paragraph{Finite Differences Baseline}
We compute Monte Carlo estimates of the gradient using central differences
\begin{equation}
    \begin{aligned}
        \frac{\partial \left(\mathbb{E}_{p(\cdot \mid \theta)}[\Phi(x)]\right)_{j}}{\partial \theta_{i}} \: \approx \:
        &\frac{1}{N_{\text{FD}}}\sum_{k=1}^{N_{\text{FD}}}\frac{[\Phi(x^{+}_{k})]_{j} - [\Phi(x_{k}^{-})]_{j}}{2\epsilon},
    \end{aligned}
\end{equation}
where samples $x^{+}_{k} \sim p(\cdot \mid \theta + \epsilon e_{i}) \text{ and } x^{-}_{k} \sim p(\cdot \mid \theta - \epsilon e_{i})$ are obtained through forward simulation. To achieve low-bias gradient estimates, we carefully tune the step size $\epsilon > 0$ to be sufficiently small while ensuring numerical stability. Similarly, we select the number of simulations $N_{\text{FD}}$ to adequately reduce estimation variance. Both hyperparameters are tuned independently for each model and parameter combination. Additional experimental details are provided in \autoref{app:exp_details}.

\subsection{Axtell's Model of Firms}

We begin our validation with AMOF, which contains eight parameters governing the beta distributions from which agent and firm characteristics are sampled. Specifically, agents sample their work-leisure preferences $\theta_i$ from $\text{Beta}(\theta_\alpha, \theta_\beta)$ and initial effort levels $e_i(0)$ from $\text{Beta}(e_\alpha, e_\beta)$. Firms sample their production scale $a_j$ from $\text{Beta}(a_\alpha, a_\beta)$ and return increase $b_j$ from $\text{Beta}(b_\alpha, b_\beta)$. We focus on the time series of mean firm output as our emergent observable $\Phi(x)$, representing a key macroeconomic indicator that emerges from the micro-level interactions. 

\begin{figure}[h!]
    \centering
    \includegraphics[width=0.8\linewidth]{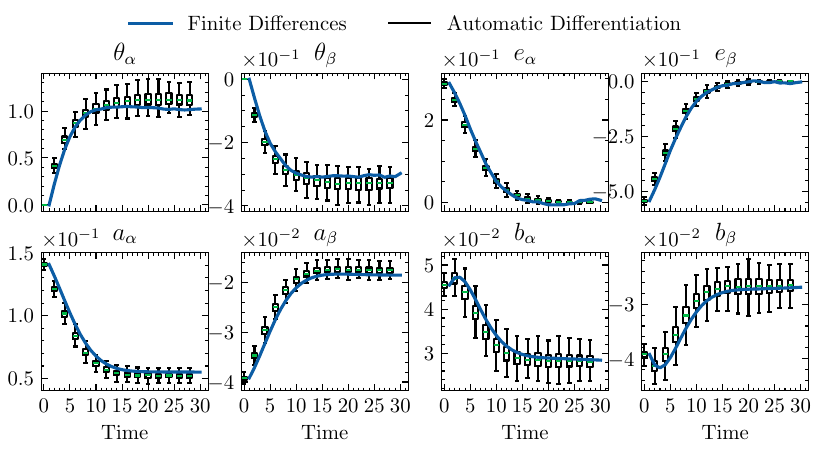}
    \caption{Validation of AD gradient estimates against FD baselines for AMOF. Each panel displays the gradient of mean firm output with respect to one of eight model parameters over time. Blue lines represent FD baselines while white box plots show AD estimates with their associated uncertainty.}
    \label{fig:axtell_grads}
\end{figure}

\autoref{fig:axtell_grads} presents the validation results, showing strong agreement between FD and AD gradient estimates across all parameters and time steps. The AD estimates (shown as white box plots) closely track the FD baselines (blue lines), demonstrating that our differentiable implementation accurately captures the gradient dynamics of this complex economic model.

\subsection{Sugarscape Model}
\label{sec:sugarscape_val}

We next validate our approach on the Sugarscape model, which presents additional complexity through its categorical parameter distributions and spatial dynamics. The model contains six parameters that define agent characteristics: metabolic rates $m_i$ and initial wealth $h_i(0)$ are sampled from beta distributions $\text{Beta}(m_\alpha, m_\beta)$ and $\text{Beta}(w_\alpha, w_\beta)$ respectively, while vision ranges are sampled from a categorical distribution $\text{Cat}(p)$. For simplicity, we restrict vision ranges to two values (one and three cells), making $p$ a two-dimensional probability vector with entries $p_{1}$ and $p_{3}$.

We examine the time series of average agent wealth as our emergent observable $\Phi(x)$, which captures the macroeconomic dynamics arising from individual foraging and spatial competition. The model is initialized with 100 agents distributed across a $50 \times 50$ grid. This scale is necessary for stable gradient estimation—smaller configurations resulted in prohibitively high variance in FD estimates, particularly for the vision range parameters $p_{1}$ and $p_{3}$, since small perturbations in these probability parameters produce negligible changes in agent composition when the total population is small.

\begin{figure}
    \centering
    \includegraphics[width=\linewidth]{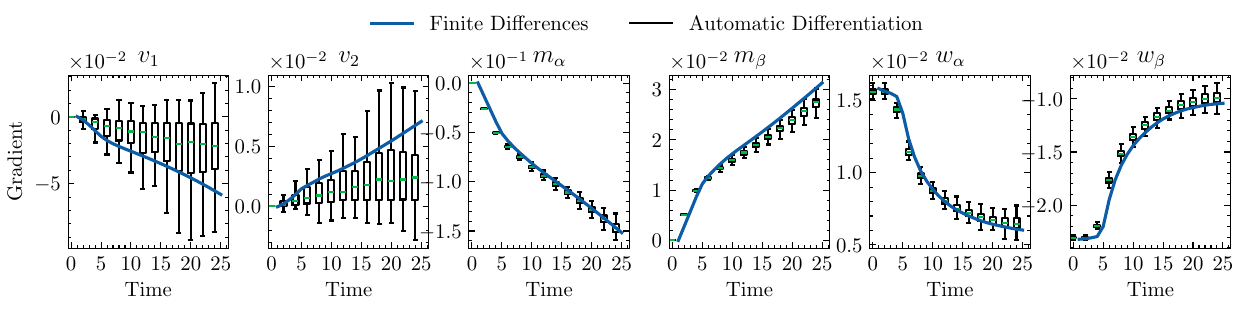}
    \caption{Validation of AD gradient estimates against FD baselines for Sugarscape parameters. Each panel displays the gradient of mean agent wealth with respect to one of six model parameters over time. Blue lines represent FD baselines while white box plots show AD estimates. Note the divergence in vision range parameters ($p_1$, $p_3$) at later time steps, reflecting the challenges of differentiating through categorical distributions in spatially-structured models.}
    \label{fig:sugarscape-gradients}
\end{figure}

\autoref{fig:sugarscape-gradients} presents our validation results. The results reveal important differences in gradient quality across parameter types. AD gradient estimates for the continuous parameters governing initial wealth and metabolism (panels corresponding to $w_\alpha$, $w_\beta$, $m_\alpha$, $m_\beta$) demonstrate excellent agreement with FD baselines throughout the simulation period.

However, the categorical vision range parameters $p_{1}$ and $p_{3}$ present greater challenges. While AD estimates initially align well with FD baselines, they exhibit increasing variance and divergence over time. This behaviour may reflect both the inherent difficulty of differentiating through categorical distributions and the relatively small magnitude of these gradients—when the true gradient signal is weak, estimation noise becomes more prominent relative to the underlying signal. Despite this increased uncertainty, the AD gradient estimates for vision parameters remain sufficiently informative to guide optimization procedures, as we demonstrate in our calibration experiments (\autoref{sec:calib}).

\subsection{SIR Model}
\label{sec:sir_gradients}

\begin{figure}
    \centering
    \includegraphics[width=0.7\linewidth]{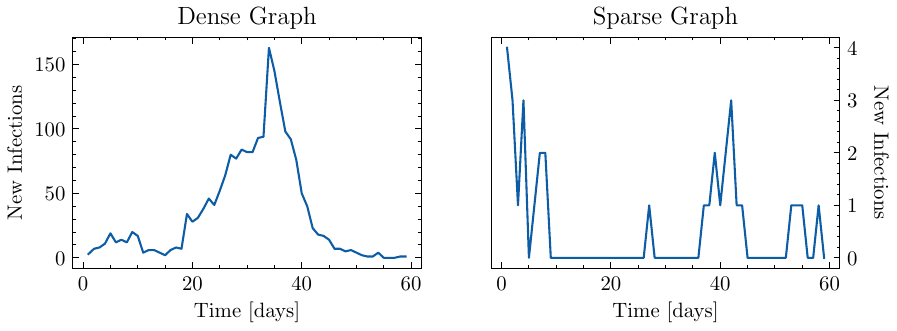}
    \caption{Infection dynamics under different contact structures. Left: complete contact graph where every agent interacts with all others. Right: sparse contact graph generated from an Erdős–Rényi random graph with edge probability $5 \times 10^{-3}$. The complete graph produces smoother dynamics that closely approximate the classical SIR ODE system.}
    \label{fig:sir-output}
\end{figure}

Finally, we validate our approach on the SIR epidemiological model, which presents unique challenges through its combination of discrete sampling procedures and differentiable control flow structures. The model contains nine parameters: three global epidemiological parameters (initial infection probability $I_{0}$, transmission rate $\beta$, and recovery rate $\gamma$), three quarantine policy parameters (start time $Q_{\text{start}}$, end time $Q_{\text{end}}$, and compliance probability $p_{Q}$), and three social distancing parameters (start time $D_{\text{start}}$, end time $D_{\text{end}}$, and reduction factor $\alpha_{D}$). The temporal parameters $Q_{\text{start}}$, $Q_{\text{end}}$, $D_{\text{start}}$, and $D_{\text{end}}$ involve differentiable control flow, while the remaining parameters affect discrete sampling procedures.

We consider a population of $N=2000$ agents with unit time steps ($\Delta t = 1$), focusing on the time series of daily infections as our emergent observable $\Phi(x)$. This metric captures the aggregate epidemiological dynamics that emerge from individual-level infection processes and policy interventions. 

We first examine the case where agents interact through a complete contact graph, meaning each agent can potentially infect any other agent at each time step (\autoref{fig:sir-output}, left panel). This configuration is particularly amenable to gradient estimation because when the contact graph is complete and both $\beta \Delta t \ll 1$ and $\gamma \Delta t \ll 1$, our ABM converges to a low-variance discretization of the classical SIR ordinary differential equation system (details in \autoref{app:sir_ode_connection}). Thus, in the case of a complete contact graph it suffices to track aggregate quantities in order to obtain meaningful estimations of the gradient. For this reason, we employ the ST estimator, which replaces Bernoulli samples with their expectation, to smoothly approximate discrete sampling procedures during the tangent pass. 

\begin{figure}[t]
    \centering
    \includegraphics[width=\linewidth]{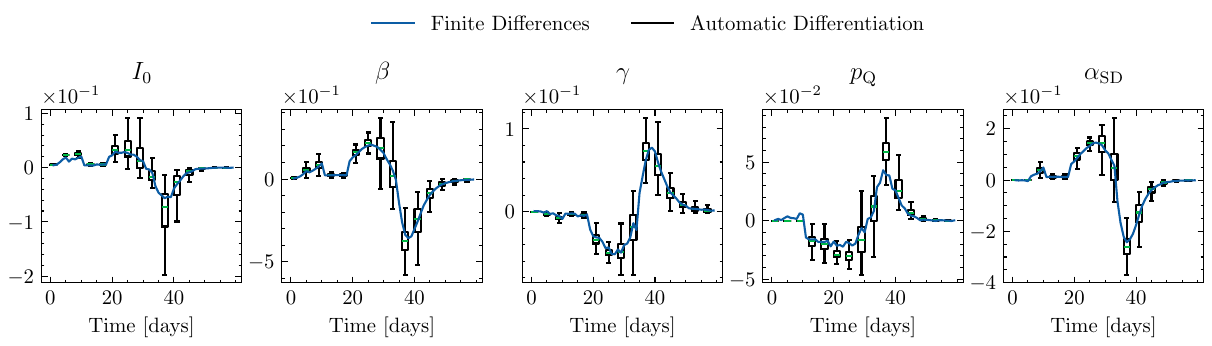}
    \caption{Validation of AD gradient estimates for core SIR parameters. Each panel displays the gradient of average daily infections with respect to one of five epidemiological and policy parameters over time. Blue lines represent FD baselines while white box plots show AD estimates.}
    \label{fig:sir_gradients_time}
\end{figure}

\autoref{fig:sir_gradients_time} demonstrates excellent agreement between AD and FD gradient estimates for the five core parameters $I_0$, $\beta$, $\gamma$, $p_Q$, and $\alpha_{SD}$. The AD estimates exhibit low variance and closely track the FD baselines throughout the simulation period, confirming the effectiveness of our differentiable implementation for standard epidemiological parameters.

\begin{figure}
    \centering
    \includegraphics[width=0.7\linewidth]{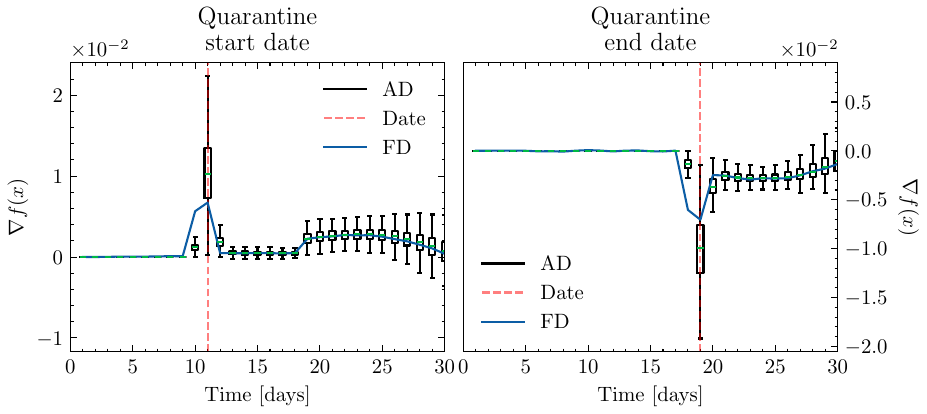}
    \caption{Validation of AD gradient estimates for quarantine timing parameters. Left panel shows gradients with respect to quarantine start time $Q_{\text{start}}$, right panel shows gradients with respect to quarantine end time $Q_{\text{end}}$. Red vertical dashed lines indicate the actual policy timing. Blue lines represent FD baselines while white box plots show AD estimates.}
    \label{fig:quarantine_gradients}
\end{figure}

The temporal policy parameters present additional complexity due to their involvement in control flow structures. \autoref{fig:quarantine_gradients} shows validation results for the quarantine timing parameters $Q_{\text{start}}$ and $Q_{\text{end}}$. The gradient estimates exhibit intuitive behaviour: they are zero before quarantine implementation, reach maximum magnitude near quarantine initiation (where infection rates change most rapidly), and remain sensitive after quarantine ends due to the lasting effects of intervention timing on post-policy case counts. For $Q_{\text{start}}$, gradients decrease to zero shortly after quarantine begins as infections consistently drop to low levels. These results validate our differentiable control flow approach from \autoref{sec:multi_branch} and demonstrate that policy timing parameters can be effectively optimized using gradient-based methods.

\subsubsection{Comparison of Estimators for Differentiable Discrete Randomness}
\label{sec:estimators_comparison}

The validation results presented above were conducted using a complete contact graph and the straight-through (ST) estimator. We now extend our analysis to compare different gradient estimators for discrete sampling procedures (introduced in \autoref{sec:discrete_randomness}) and evaluate their performance under both complete and sparse network topologies. This comparison is crucial for understanding when each estimator is appropriate and highlights the fundamental trade-offs between bias, variance, and computational cost across different model structures.

We evaluate four different gradient estimators:
\begin{itemize}
    \item The ST gradient estimator, which introduces significant bias by using the identity function as a surrogate during the tangent pass.
    \item The GS gradient estimator with a finely tuned temperature $\tau = 0.1$ aiming to  trade-off bias and variance.
    \item The \texttt{StochasticAD.jl} gradient estimator, which provides unbiased estimates at the cost of significant variance. We use ten samples per estimation, making this estimator ten times more expensive than the ST and GS gradient estimators.
    \item The smoothed \texttt{StochasticAD.jl} gradient estimator. This estimator has comparatively lower variance than its vanilla counterpart, but introduces bias.
\end{itemize}

We focus our comparison on the initial infection probability $I_0$ and transmission rate $\beta$, as these parameters are invoked repeatedly in Bernoulli sampling operations throughout each simulation, making gradient estimation particularly sensitive to the choice of estimator.

\begin{figure}[!htb]
    \centering
    \begin{subfigure}[b]{0.9\linewidth}
        \centering
        \includegraphics[width=\linewidth]{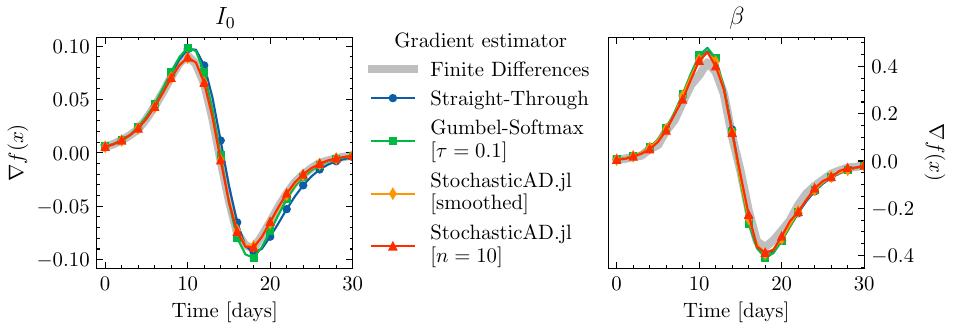}
        \caption{\textbf{Dense Connectivity (Complete Graph)}}
        \label{fig:sir_gradients_dense}
    \end{subfigure}
    
    \vspace{1em}
    
    \begin{subfigure}[b]{0.9\linewidth}
        \centering
        \includegraphics[width=\linewidth]{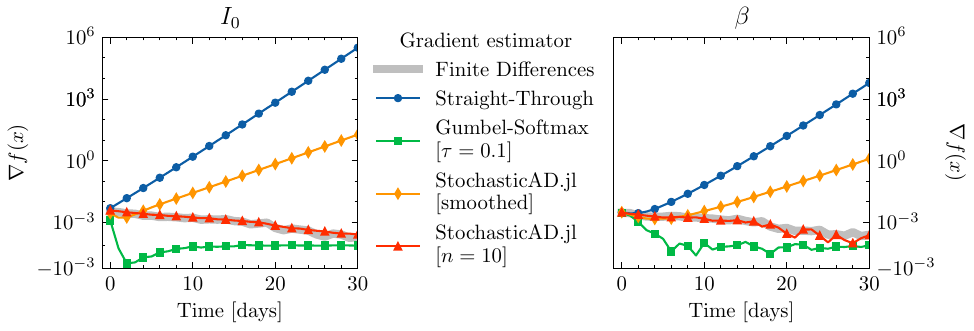}
        \caption{\textbf{Sparse Connectivity (Erdős–Rényi Graph, $p = 5 \times 10^{-3}$)}}
        \label{fig:sir_gradients_sparse}
    \end{subfigure}
    
    \caption{Gradient estimator comparison under different connectivity patterns. \textbf{Top:} Under complete connectivity, each panel shows gradients of daily infections with respect to initial infection probability $I_0$ (left) and transmission rate $\beta$ (right). All estimators perform well due to the smooth, ODE-like dynamics. \textbf{Bottom:} Under sparse connectivity, the plot shows error distributions for each estimator relative to FD baselines. Note the logarithmic y-axis scale and the dramatic failure of biased estimators.}
    \label{fig:sir_gradients_combined}
\end{figure}

\paragraph{Complete contact network} Under the complete contact graph configuration, all gradient estimators perform exceptionally well, closely matching FD baselines (\autoref{fig:sir_gradients_dense}). This result is expected given our previous validation showing that the ST estimator already achieved excellent agreement with FD baselines in this setting. The strong performance across all methods reflects the underlying mathematical structure: when every agent can contact every other agent, our discrete agent-based model closely approximates the classical SIR ordinary differential equation system. In this regime, the exchange of gradient and expectation operations is valid, eliminating bias and allowing even the nominally biased ST estimator to perform as well as more sophisticated alternatives.

\paragraph{Sparse contact network results.} The estimator comparison reveals stark differences when we transition to a sparse contact network generated from an Erdős–Rényi random graph with edge probability $5 \times 10^{-3}$ (\autoref{fig:sir-output}, right panel). Under this configuration, agents interact with only 0.5\% of the population on average at each time step, breaking the connection to smooth ODE dynamics and making aggregate-level approximations insufficient for accurate gradient estimation.

\autoref{fig:sir_gradients_sparse} demonstrates the dramatic performance differences under sparse connectivity. The ST estimator fails catastrophically, producing exploding gradients that diverge by orders of magnitude from the true values (note the logarithmic y-axis). The GS estimator exhibits severe underestimation and vanishing gradients. The smoothed StochasticAD.jl estimator initially tracks FD baselines but diverges substantially after ten time steps as accumulated bias compounds over time.

Only the standard StochasticAD.jl estimator maintains consistent agreement with FD baselines throughout the simulation, validating its theoretical unbiasedness guarantees. These results underscore a critical insight: for highly non-linear ABMs, the bias introduced by smoothed estimators can be fatal to gradient-based optimization. While StochasticAD.jl suffers from high variance and thus we need additional computational resources, it may be our only option when calibrating ABMs at certain regimes.

\subsection{Sensitivity Analysis via Automatic Differentiation}
\label{sec:sa}

Before proceeding to calibration experiments, we discuss how automatic differentiation (AD) enables efficient sensitivity analysis for agent-based models (ABMs). Sensitivity analysis examines how uncertainty and variation in model outputs can be attributed to individual model parameters, helping researchers identify which agent attributes or interaction rules most significantly impact emergent phenomena. This understanding is crucial for comprehending complex system mechanisms and designing effective policy interventions.

\paragraph{Limitations of classical methods} Traditional sensitivity analysis approaches face significant computational challenges when applied to ABMs. Local methods, such as one-at-a-time (OAT) perturbations and FD-based derivative estimates, require sequential parameter perturbations, leading to a computational cost that scales linearly with the number of parameters, $\mathcal{O}(d)$, for $d$ parameters \citep{ten_broeke_which_2016}. FD methods also suffer from numerical instabilities due to the challenge of selecting appropriate step sizes, particularly in complex, non-linear systems like ABMs. Global sensitivity analysis methods, such as Sobol indices and Shapley value-based approaches, quantify parameter contributions to output variance across the entire input space, requiring $N(d+2)$ model evaluations, where $N$ ranges from hundreds to thousands depending on model complexity \citep{saltelli_making_2002}. These methods are computationally prohibitive for large-scale ABMs, despite their ability to capture non-linear interactions and global effects.

\paragraph{Advantages of AD-Based sensitivity analysis} Our differentiable ABM framework exploits forward and reverse mode AD to calculate exact first-order local sensitivities (Jacobians) across many parameters in a single execution, cutting complexity from $\mathcal{O}(d)$ to essentially $\mathcal{O}(1)$ runs \citep{quera-bofarull_dont_2023}. For scalar outputs, reverse-mode AD handles all parameter gradients in one backward pass, making it practical to analyze ABMs with thousands of parameters. Unlike FD-based approaches, AD sidesteps numerical biases and the headache of step-size selection while delivering precise derivatives. While global methods like Sobol indices reveal parameter importance across input spaces, our AD approach shines for local sensitivity analysis---particularly useful for parameter tuning, optimization, and equilibrium analysis. These local sensitivities can actually inform global analysis by directing efficient sampling strategies \citep{rakovec_distributed_2014, naumann-woleske_exploration_2024}.

\paragraph{Temporal sensitivity insights} AD-based sensitivity analysis efficiently captures how parameter influences evolve over time in ABMs. The policy timing gradients for the SIR model (\autoref{fig:quarantine_gradients}) identify precisely when quarantine interventions have maximum impact on infection dynamics, providing policymakers with actionable guidance on intervention timing. The AMOF gradient estimates (\autoref{fig:axtell_grads}) reveal how initial effort distribution parameters $e_{\alpha}$ and $e_{\beta}$ decay to zero over time, indicating when the economic system effectively ``forgets'' agents' initial conditions. This temporal decay pattern distinguishes between persistent and transient initial condition effects, which proves essential for understanding path dependence---some starting conditions influence the entire simulation, while others are overwhelmed by endogenous dynamics. While finite difference methods can capture temporal patterns, they require multiple evaluations per parameter and introduce numerical errors. AD provides superior efficiency and accuracy for analyzing critical transitions, periods of maximum parameter influence, and the temporal decay of initial condition effects.


\section{Calibration of Differentiable Agent-Based Models}

\label{sec:calib}

In this section, we demonstrate how access to gradients can facilitate fast Bayesian calibration of ABMs. In particular, we calibrate each of the ABMs introduced in \autoref{sec:abms} via a variational approach that leverages the differentiable implementations developed in \autoref{sec:diff_abms}. Before presenting our experimental results in \autoref{sec:calib_exps}, we first provide necessary background on model calibration next.

\subsection{Calibration via Generalised Variational Inference}
\label{sec:calib_intro}
Given an ABM defined as a stochastic map $f: \Theta \to \mathcal{X}$ with parameters $\theta \in \Theta$, the broad goal of calibration procedures is to find parameter values that are consistent with observed emergent data $y \in \mathbb{R}^n$. Here, $y$ represents empirically observed summary statistics that we assume were generated by some true but unknown process, which we seek to approximate with our ABM. In particular, Bayesian calibration procedures aim to produce a posterior distribution $p(\theta \mid y)$ over $\Theta$ that provides natural uncertainty quantification over model parameters. 

Traditional Bayesian calibration methods face significant challenges when applied to ABMs. Beyond the usual computational difficulties of intractable likelihoods and high-dimensional parameter spaces, ABM calibration presents a fundamental limitation: any ABM constructed to study a complex real-world system is a limited representation of the true data generation process and so will likely be \textit{misspecified} to some extent. Traditional Bayesian inference can be sensitive to such misspecification, potentially leading to overconfident posterior distributions that do not adequately reflect model uncertainty \citep{cannon_investigating_2022}.

Generalised Bayesian methods address this challenge by replacing the potentially misspecified likelihood with a more flexible loss function while maintaining principled uncertainty quantification \citep{knoblauch_optimization-centric_2022}. These methods target a generalised posterior distribution over $\Theta$ of the form
\begin{equation}
    p(\theta \mid y) \propto \exp(-\ell(\theta, y)) \; p(\theta),
\end{equation}
where $\ell(\theta, y)$ is a suitable loss function \footnote{Taking $-\ell(\theta,y) = \log p(y|\theta)$ recovers classical Bayesian inference.} capturing the dissimilarity between observed data $y$ and model-generated summary statistics $\Phi(f(\theta))$. The choice of loss function $\ell$ allows practitioners to incorporate domain knowledge about which aspects of the data are most important to match.

While various computational approaches exist for approximating this generalised posterior, we focus on generalised variational inference (GVI) \citep{knoblauch_optimization-centric_2022, cherief-abdellatif_mmd-bayes_2020}, which uses a variational approximation strategy.

The generalised posterior may be approximated by solving a variational optimisation problem in which one minimises the KL-divergence between a variational distribution $q_{\phi}(\theta)$ with parameters $\phi \in \Phi$, and the generalised posterior,
\begin{equation}\label{eq:GVI}
    \min_{\phi} \left\{ \mathcal{L}(\phi) := \mathbb{E}_{\theta \sim q_\phi(\theta)}\left[\ell(\theta,y) + \log \frac{q_\phi(\theta)}{p(\theta)}\right] \right\}.
\end{equation}

Note that the first term in \autoref{eq:GVI} assesses the quality of model fit, whilst the second may be interpreted as regularisation term that encourages the variational posterior $q_{\phi}$ to stay close to the prior. Using a score-based gradient estimator for $\mathcal{L}$, one may train the variational posterior through stochastic gradient descent on the variational parameters $\phi$. However, score-based estimators do not rely on differentiability of the ABM and as a result can suffer high variance \citep{mohamed_monte_2020}. Instead, we derive an alternative gradient estimator that exploits differentiability. Before proceeding, we require assumptions on the variational family and the form of the loss function $\ell$.
\begin{assumption}
    \label{ass:calib_ass}
    We assume that the following conditions hold:
    \begin{enumerate}[label=(\roman*)]
        \item The prior $p(\theta)$ is differentiable with respect to $\theta$.
        \item  $\ell(\theta, y) := \mathbb{E}_{x \sim p(\cdot \mid\theta)}[h(x, y)]$ for some function $h$ which is differentiable in its first argument.
        \item For all $\phi \in \Phi$, $q_{\phi}(\cdot)$ is differentiable and tractable to evaluate.
        \item Sampling from $q_{\phi}$ is reparameteriasble. That is, there exists a function $g$, which is differentiable in its first argument, and a random variable $Z$, such that for all $\phi \in \Phi$:
        \begin{equation}
            z \sim Z \implies g(\phi, z) \sim q_{\phi}.
        \end{equation}
    \end{enumerate}
\end{assumption}
By employing the reparameterisation trick, we may now rewrite $\mathcal{L}(\phi)$ as,
\begin{equation}
         \mathcal{L}(\phi) = \mathbb{E}_{z \sim Z}\left[\mathbb{E}_{x \sim p(\cdot \mid g(\phi, z))}[h(x, y)] + \log \frac{q_\phi(g(\phi, z))}{p(g(\phi, z))}\right].
\end{equation}
With $\mathcal{L}(\phi)$ in this form, we may express the gradient $\nabla_{\phi}\mathcal{L}(\phi)$ as follows,
\begin{equation}
    \nabla_{\phi}\mathcal{L}(\phi) = \mathbb{E}_{z \sim Z}\left[\nabla_{\phi}g(\phi, z)\cdot\nabla_{\theta}\mathbb{E}_{x \sim p(\cdot \mid \theta)}[h(x, y)]\big\vert_{\theta = g(\phi, z)} + \nabla_{\phi}\log \frac{q_\phi(g(\phi, z))}{p(g(\phi, z))}\right].
\end{equation}
Therefore, we may estimate $\nabla_{\phi}\mathcal{L}(\phi)$ via the Monte Carlo sum,
\begin{equation}
\label{eq:calib_grads}
    \nabla_{\phi}\mathcal{L}(\phi) \approx \frac{1}{B}\sum^{B}_{b=1}\nabla_{\phi}g(\phi, z^{(b)})\cdot \eta^{(b)} + \nabla_{\phi}\log\frac{q_{\phi}(g(\phi, z^{(b)}))}{p(g(\phi, z^{(b)}))},
\end{equation}
where $z^{(b)} \sim Z$, $B$ is the batch size, and $\eta^{(b)}$ is the approximation,
\begin{equation}
    \eta^{(b)} \approx \nabla_{\theta} \mathbb{E}_{x \sim p(\cdot \mid \theta)}[h(x, y)]\big\vert_{\theta = g(\phi, z^{(b)})},
\end{equation}
computed using AD. Note that forward-simulation of the ABM with the parameters $\theta = g(\phi, z^{(b)})$ is required to compute $\eta^{(b)}$ so that the primal and tangent passes can be performed. Using \autoref{eq:calib_grads}, we may solve  \autoref{eq:GVI} via stochastic gradient descent in order to compute a variational approximation of the generalised posterior. We refer to \autoref{eq:calib_grads} as the \emph{pathwise} gradient estimator for $\nabla_{\phi}\mathcal{L}(\phi)$.

\paragraph{Normalizing flows} The choice of a variational family $q_\phi$ is an important consideration. Not only is our ability to recover the true posterior dependent on the expressiveness of the selected family, but the above derivation relies on the ability to reparameterise the sampling process. 
A convenient choice of variational family, given both desiderata, are normalizing flows. These transform a simple base distribution, such as a standard normal, into a more complex distribution through a series of invertible transformations,
\begin{equation}
    q_\phi(\theta) = T_\phi(z), \quad z \sim \mathcal{N}(0, I),
\end{equation}
where $T_\phi$ is a composition of invertible neural networks. This allows us to capture complex dependencies between parameters while maintaining computational tractability of the density $q_{\phi}$ through the change of variables formula
\begin{equation}
    \log q_\phi(\theta) = \log p_{Z}(z) - \log \left|\det\frac{\partial T_\phi}{\partial z}\right|,
\end{equation}
where $p_{Z}$ is the base density associated with $Z$. 

\paragraph{Constrained parameter domains through bijector transformations} An important practical consideration when using normalizing flows for parameter inference is ensuring that samples from the flow remain within the valid parameter domain. While the flow can theoretically sample from the entire real space, ABM parameters often have natural constraints (e.g., positive rates, bounded probabilities). To address this, we employ bijector transformations as the final layer of the flow architecture \citep{dillon_tensorflow_2017, kucukelbir_automatic_2017}. These bijectors automatically map from the unconstrained flow domain to the constrained prior support, ensuring that all samples $\theta = T_\phi(z)$ satisfy the required parameter constraints. This approach leverages AD libraries such as \texttt{Bijectors.jl} \citep{fjelde_bijectorsjl_2020} to handle the necessary Jacobian computations for maintaining proper probability densities under the constraint transformations.

A wide range of flow architectures have been proposed in the literature, each with their own strengths and weaknesses. We refer the curious reader to the excellent monograph of \cite{papamakarios_normalizing_2021}. In our experiments, we use masked autoregressive flows \citep{papamakarios_masked_2017}, but could have used any suitable alternative. Full details regarding the flow architecture used in each calibration experiment are given in \autoref{app:exp_details}.

\subsection{Hybrid Automatic Differentiation Strategy}

The choice of automatic differentiation mode has significant computational implications for our variational inference approach. In calibration tasks, the objective function is typically a scalar loss, making reverse-mode AD theoretically attractive due to its favourable scaling with output dimension. However, the memory requirements of reverse-mode AD pose substantial challenges for ABMs. Unlike forward-mode AD, which computes gradients concurrently with the forward pass, reverse-mode AD necessitates storing the entire computational graph during simulation to enable subsequent backpropagation. For ABMs featuring large agent populations that evolve over many time steps, this storage requirement leads to prohibitive memory consumption that scales with both population size and simulation length \citep{quera-bofarull_challenges_2023}.

Forward-mode AD circumvents these memory constraints entirely. Since the tangent pass executes concurrently with the primal computation, no computational graph needs to be stored—gradients are computed on-the-fly and the intermediate computations can be discarded immediately. This maintains constant memory overhead regardless of simulation length.

How can we harvest the advantageous space complexity of forward-mode AD with the favourable time complexity of reverse-mode AD? After all, normalizing flows typically contain thousands of parameters, making forward-mode differentiation through the flow computationally prohibitive. The solution lies in recognizing the compositional structure of our gradient computation from \autoref{eq:calib_grads}. Recall that we require gradients of the form $\nabla_{\phi}g(\phi, z^{(b)}) \cdot \eta^{(b)}$, where $g(\phi, z) = T_\phi(z)$ represents sampling from the normalizing flow and $\eta^{(b)} = \nabla_{\theta} \mathbb{E}_{x \sim p(\cdot \mid \theta)}[h(x, y)]$ captures the ABM's response to parameter changes. We can decompose this computation by applying reverse-mode AD to differentiate through the flow transformation $T_\phi$, yielding the Jacobian $\nabla_{\phi}T_\phi(z^{(b)})$, and forward-mode AD to compute the ABM gradients $\eta^{(b)}$. The full gradient is then obtained through matrix-vector multiplication of these two components. This hybrid approach ensures the constant memory overhead of forward-mode AD for ABM simulation while preserving the computational efficiency of reverse-mode differentiation through the high-dimensional flow parameters.

\subsection{Calibration Results}
\label{sec:calib_exps}

Next, we apply the calibration procedure outlined in Section \ref{sec:calib_intro} to each of the differentiable ABMs implemented in Section \ref{sec:diff_abms}. In each case, we calibrate to synthetic data generated by forward-simulating the ABM using a set of ground-truth parameters. As a benchmark, we adopt variational posteriors trained using the score-based gradient estimator \texttt{VarGrad} \citep{richter_vargrad_2020}. We select \texttt{VarGrad} as our primary comparison due to its demonstrated superiority over the standard \texttt{REINFORCE} score estimator \citep{williams_simple_1992}, providing a more challenging and representative baseline for evaluating our pathwise gradient approach. This choice ensures that any performance gains observed for the pathwise estimator are meaningful when compared against the current state-of-the-art in score-based methods. We chose $\ell$ to be the maximum mean discrepancy (MMD) loss due to its theoretical advantages over other losses (see \cite{knoblauch_optimization-centric_2022, gretton_kernel_2012}). Further experimental details describing ground truth parameters, prior distributions used, flow architectures, and training hyperparameters are shown in \autoref{app:exp_details}.

\subsubsection{Axtell's Model of Firms}

\begin{figure}
    \centering
    \includegraphics[width=0.95\linewidth]{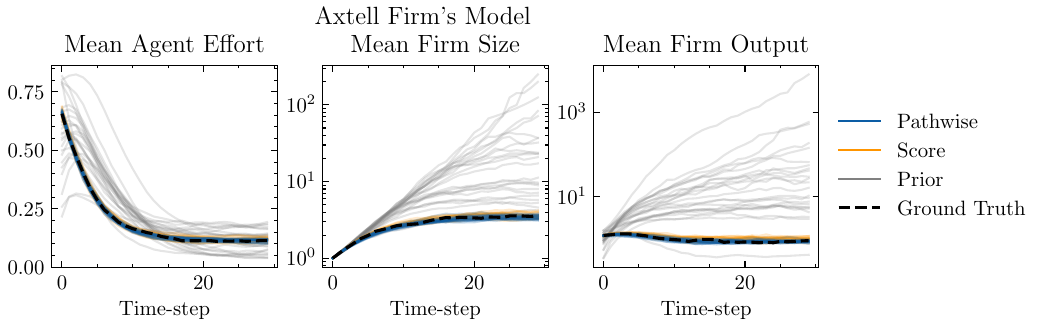}
    \caption{Calibration of AMOF. Both panels show time-series generated using parameters sampled from the prior (grey), the variational posterior learned using the pathwise gradient estimator (blue), and the variational posterior learned using a score-based gradient estimator (orange). The ground truth time-series data used for calibration is denoted by dashed lines.}
    \label{fig:axtell_data_plots}
\end{figure}

For AMOF, we calibrate to the three-dimensional time-series tracking mean agent effort, mean firm size and mean firm output. \autoref{fig:axtell_data_plots} displays time-series data generated using parameters sampled from a variational posterior trained with pathwise gradients, and a variational posterior trained using \texttt{VarGrad}. Both posteriors produce time-series that closely match the ground-truth data, significantly improving upon time-series generated via the prior. \autoref{fig:axtell_elbo} shows the loss per gradient descent iteration for both gradient estimators. We observe that the calibration with the pathwise estimator converges faster than the score-based one, also showing greater stability likely due to the lower variance. \autoref{fig:axtell_posterior} shows the inferred generalized posterior distribution, demonstrating that both methods assign high probability mass to the true parameters.

\subsubsection{Sugarscape Model}

For Sugarscape, we calibrate to the two-dimensional time-series tracking mean agent holdings and the fraction of agents that are alive. We calibrate the vision probability vector $p$ and the wealth and metabolism parameters $w_{\alpha}$, $w_{\beta}$, $m_{\alpha}$, and $m_{\beta}$. To ensure that vision parameters play a meaningful role in calibration, we restrict the vision possibilities to two values: vision range 1 and vision range 6, as the model outputs are not highly sensitive to vision values.

\begin{figure}[h]
    \centering
    \includegraphics[width=0.9\linewidth]{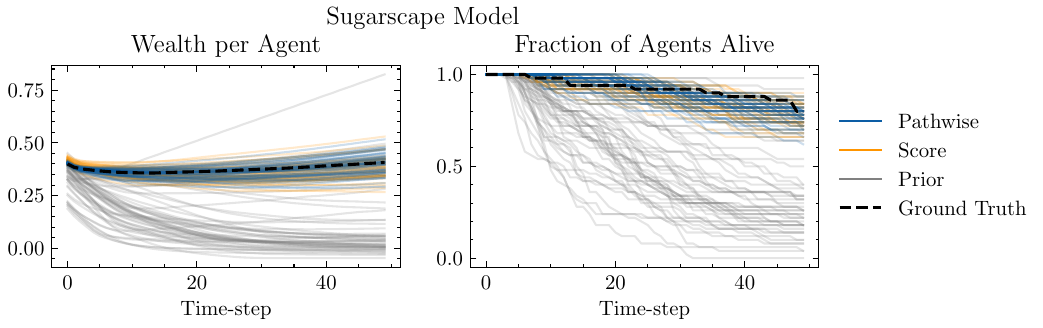}
    \caption{Calibration of all free parameters ($v_1$, $v_2$, $w_{\alpha}, w_{\beta}, m_{\alpha}, m_{\beta}$) in Sugarscape. Time-series generated by forward-simulating parameters drawn from the prior (grey), pathwise gradient posterior (blue), and \texttt{VarGrad} posterior (orange). Ground-truth calibration data shown as black dashed lines.}
    \label{fig:time-series-sugar}
\end{figure}

\autoref{fig:time-series-sugar} shows that the posterior predictive distributions from both pathwise and \texttt{VarGrad} estimators correctly encompass the ground-truth data, indicating successful calibration. The corresponding posterior distribution shown in \autoref{fig:sugarscape_posterior} demonstrates that both gradient estimators correctly assign high probability mass to the true parameter values. Our gradient validation in \autoref{sec:sugarscape_val} revealed that while the continuous parameters ($w_{\alpha}$, $w_{\beta}$, $m_{\alpha}$, $m_{\beta}$) showed excellent agreement with finite difference baselines, the categorical vision range parameters ($p_1$, $p_6$) exhibited increasing variance and divergence over time. However, these successful calibration results demonstrate that despite the finite difference discrepancies, our AD gradients for vision parameters remain sufficiently informative for optimization purposes. This confirms that our differentiable implementation can effectively calibrate both continuous and categorical parameters in the Sugarscape model.

\subsubsection{SIR Model}

For the SIR model, we calibrate to a two-dimensional time series tracking daily infections and recoveries. We sample a contact graph from the Erdős–Rényi random graph model with $N=2000$ agents and edge probability $10^{-2}$, which corresponds to a moderately connected graph where we expect the GS estimator (\autoref{eq:gsoftmax}) for discrete randomness to perform well (\autoref{sec:estimators_comparison}).

\begin{figure}[h]
    \centering
    \includegraphics[width=0.9\linewidth]{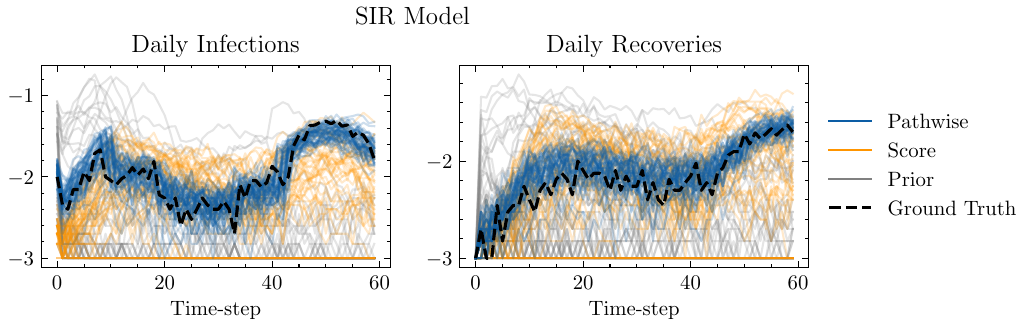}
    \caption{Calibration of the SIR model. Both panels show time series generated via forward simulation of parameters sampled from the prior (grey), a variational posterior trained with pathwise gradients (blue), and a variational posterior trained with \texttt{VarGrad} (orange). Ground truth data is shown as dashed black lines.}
    \label{fig:sir-time-series}
\end{figure}

\begin{figure}[h]
    \centering
    \begin{subfigure}[b]{0.32\textwidth}
        \centering
        \includegraphics[width=\textwidth]{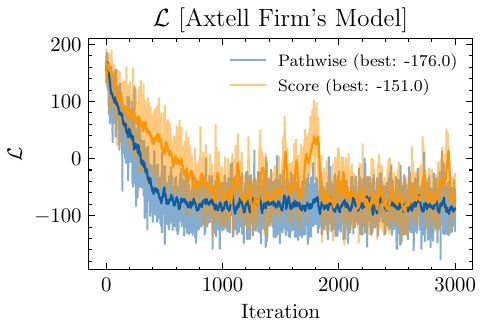}
        \caption{Axtell Firms model}
        \label{fig:axtell_elbo}
    \end{subfigure}
    \hfill
    \begin{subfigure}[b]{0.32\textwidth}
        \centering
        \includegraphics[width=\textwidth]{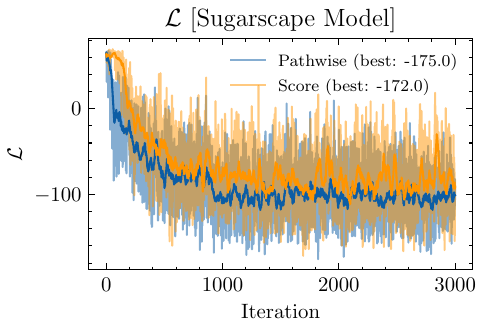}
        \caption{Sugarscape model}
        \label{fig:sugarscape_elbo}
    \end{subfigure}
    \hfill
    \begin{subfigure}[b]{0.32\textwidth}
        \centering
        \includegraphics[width=\textwidth]{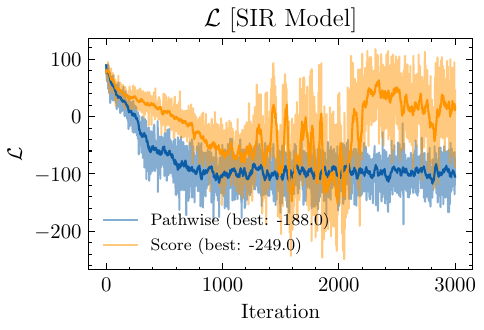}
        \caption{SIR model}
        \label{fig:sir_elbo}
    \end{subfigure}
    \caption{Loss per gradient descent iteration when training variational posteriors for each model using pathwise gradients (blue) and score-based gradients (orange). Faint lines show actual loss values at each iteration, while solid lines show a moving average.}
    \label{fig:all_elbos}
\end{figure}

\begin{figure}[h]
    \centering
    \begin{subfigure}[b]{0.32\textwidth}
        \centering
        \includegraphics[width=\textwidth]{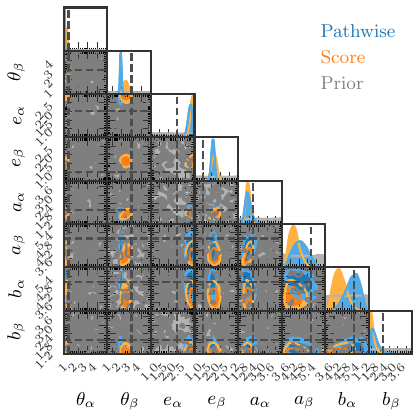}
        \caption{Axtell model of Firms}
        \label{fig:axtell_posterior}
    \end{subfigure}
    \hfill
    \begin{subfigure}[b]{0.32\textwidth}
        \centering
        \includegraphics[width=\textwidth]{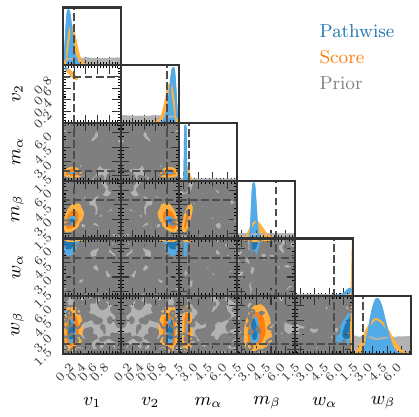}
        \caption{Sugarscape model}
        \label{fig:sugarscape_posterior}
    \end{subfigure}
    \hfill
    \begin{subfigure}[b]{0.32\textwidth}
        \centering
        \includegraphics[width=\textwidth]{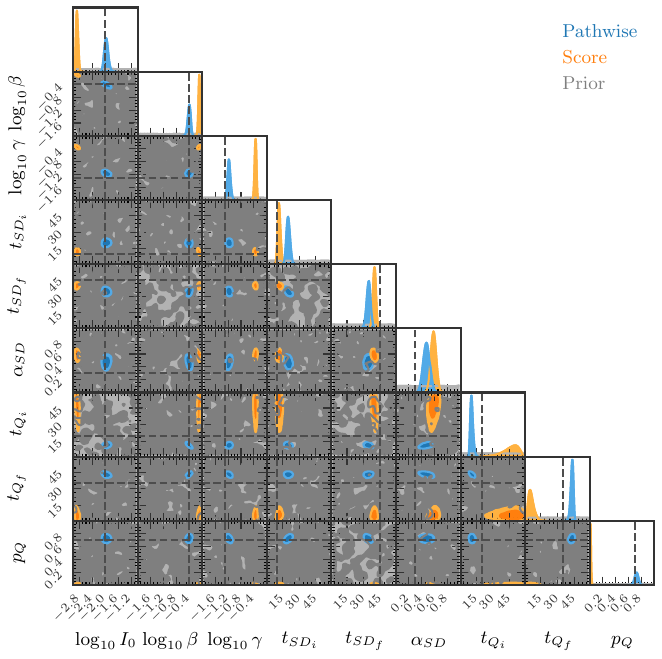}
        \caption{SIR model}
        \label{fig:sir_posterior}
    \end{subfigure}
    \caption{Corner plots showing prior distributions and trained posterior distributions using pathwise and score gradients for all three ABM models.}
    \label{fig:all_posteriors}
\end{figure}

\autoref{fig:sir-time-series} displays time series generated via variational posteriors learned using different gradient estimators. The variational posterior trained using pathwise gradients produces time series that better represent the ground truth data compared to those trained with \texttt{VarGrad}. \autoref{fig:sir_posterior} shows that the pathwise gradient successfully recovers the true parameters, assigning high probability mass to the ground truth values, while the score-based estimator fails to identify a few parameters correctly. This superior performance is reflected in the training dynamics shown in \autoref{fig:sir_elbo}, where \texttt{VarGrad} exhibits initially slower loss decrease and greater variance at later training stages, achieving a less optimal final value. This confirms our expectation that score-based methods perform worse in higher-dimensional parameter spaces than their pathwise counterparts.

\subsection{Summary of Results}

Across all the tested ABMs, pathwise gradient estimators perform no worse than \texttt{VarGrad}, a state-of-the-art score-based gradient estimator. This represents a significant achievement: obtaining informative pathwise gradients from ABMs is inherently challenging due to their discrete state spaces and extensive use of discrete randomness, making them traditionally non-differentiable. Our results demonstrate that pathwise methods can successfully extract meaningful gradient information even from these fundamentally discrete and stochastic systems. For the Axtell and Sugarscape models, pathwise gradients achieve comparable calibration performance to \texttt{VarGrad}, validating the correctness of our differentiable implementation and proving that these traditionally non-differentiable models can indeed be made amenable to gradient-based optimization.

However, the true advantage of pathwise methods becomes apparent in higher-dimensional parameter spaces. Our SIR model experiments with nine parameters reveal a clear performance gap where pathwise methods provide sufficiently informative gradients while score-based methods fail. This advantage stems from fundamental properties of score-based estimators: their variance depends heavily on both the variance of the ABM outputs and the parameter dimensionality. While the use of control variates (baselines) can partially mitigate variance issues, our experiments demonstrate this is insufficient in our setting. For more rigorous discussion on the limitations of score-based methods, we refer the reader to Section 4.3 of \cite{mohamed_monte_2020}.

As parameter spaces grow larger—which is typical in realistic ABMs with heterogeneous agent populations, complex behavioural rules, detailed environmental parameters, and increasingly, neural network components for agent decision-making—we expect pathwise methods to become not just advantageous but critical. In settings with neural-based components embedded within ABMs, pathwise gradients may become the only viable approach for end-to-end calibration of these hybrid systems. The favourable scaling of pathwise gradients with parameter dimension, combined with their robustness to model stochasticity, positions them as the essential tool for high-dimensional ABM calibration. 
\section{Relevant Literature}
\label{sec:lit}
\paragraph{Differentiable ABMs.} To the best of our knowledge, the study of differentiable agent-based modelling was initiated by \cite{andelfinger_differentiable_2021}, who replaced discontinuous operations with smooth surrogates in both the primal and tangent pass so that AD could be applied to a range of traffic and epidemiological simulators. The GS trick was first applied in the context of agent-based modelling by \cite{chopra_differentiable_2023}, who employed a combination of tensorisation and AD to construct a large-scale epidemiological model that could be calibrated using feedforward neural networks. \cite{andelfinger_towards_2023} employed a similar calibration procedure to traffic simulators. The use of generalised variational inference to calibrate differentiable agent-based models was introduced by \cite{quera-bofarull_bayesian_2023}, whilst the hybrid use of forward and reverse-mode AD during calibration was introduced by \cite{quera-bofarull_challenges_2023, dyer_gradient-assisted_2023}. Since then, the use of differentiable ABMs has extended to include more complex agent architectures, such as Large-Language-Model-based agents \citep{chopra_limits_2024}. Related work has also explored building differentiable surrogates of ABMs to reproduce their microstructure \citep{cozzi_learning_2025} and rendering the likelihood tractable for variational inference \citep{monti_learning_2023, lenti_variational_2024}.

\paragraph{Differentiable Programming.} For a general overview of AD, we refer the reader to the excellent survey of \cite{baydin_automatic_2018}. The GS trick was first introduced by \cite{maddison_concrete_2016} and has since been applied to a wide range of machine learning problems \citep{jang_categorical_2016, kusner_gans_2016, shen_variational_2021}. The ST gradient estimator has a long history tracing back to \cite{hinton_neural_2012}, and was popularised by the seminal work of \cite{bengio_estimating_2013} focused on stochastic neuron layers. The reparameterisation trick was popularised through its application to variational autoencoders by \cite{kingma_auto-encoding_2022}. A variety of smooth surrogates for control flow statements have been proposed; the thesis of \cite{petersen_learning_2022} provides a good overview. Likewise, we refer the reader to the excellent monographs of \cite{blondel_elements_2024} and \cite{niculae_discrete_2024} for further discussion and development of smooth relaxations for the $\argmax$ operator. Stochastic AD was introduced by \cite{arya_automatic_2022}, and takes heavy inspiration from smoothed perturbation analysis, which was predominantly developed by the discrete-event systems community. An overview of stochastic gradient approximation schemes from the discrete-event systems perspective is given by \cite{fu_optimization_1997}, while \cite{mohamed_monte_2020} provide an overview of Monte Carlo gradient estimation from the machine learning viewpoint.

\paragraph{Calibration.} A detailed overview of Bayesian calibration in the context of agent-based modelling is provided by \cite{dyer_black-box_2024}. Calibration has intimate connections to the field of simulation-based inference \citep{cranmer_frontier_2020}, wherein one attempts to infer parameters for statistical models implicitly defined through simulators and compute programs. The use of variational inference to calibrate black-box models was first proposed by \cite{ranganath_black_2014} and further developed in the SBI context by \cite{glockler_variational_2022}. We recommend the monograph of \cite{knoblauch_optimization-centric_2022} for a thorough overview of generalised variational inference. To the best of our knowledge, the first method for simulator calibration based on generalised Bayesian inference was proposed by \cite{gao_generalized_2023}.  \\

\section{Discussion}
\label{sec:discussion}

\subsection{Summary of Contributions}

We have demonstrated that ABMs can be made amenable to AD while preserving their inherently discrete nature, enabling path-wise gradient-based inference and optimization. Our experimental results across three distinct models—the Axtell Model of Firms, Sugarscape, and an agent-based SIR model—reveal important insights about both practical implementation and theoretical implications of differentiable ABMs.

\paragraph{Gradient validation and estimator performance} We thoroughly validated the gradients of differentiable ABMs against finite differences, confirming that our framework produces accurate gradients across different models and parameters. This accuracy is particularly evident in well-understood cases, such as the SIR model on complete graphs that match ODE predictions.

Our analysis reveals that the effectiveness of different gradient estimators for discrete randomness depends critically on how much the ABM deviates from mean field solutions. The Straight-Through estimator performs adequately when mean field approximations hold, but fails when ABMs exhibit strong nonlinear dynamics that violate mean field assumptions, as observed in disease spread on sparse networks. The unbiased \texttt{StochasticAD.jl} estimator, based on stochastic perturbation analysis, proves most robust across different scenarios by properly accounting for these deviations, though further work is needed to reduce its computational cost.


\paragraph{Sensitivity analysis} We computed sensitivities across epidemiological parameters, agent characteristics, and policy timing using automatic differentiation. This allows us to obtain local sensitivity information directly within the ABM, including through differentiable control flow for policy interventions such as quarantine and social distancing. Our approach provides gradients not only with respect to compliance rates but also intervention start and end dates, offering valuable insights for designing policies that mitigate disease spread while balancing social and economic costs.

\paragraph{Calibration pipeline and generalized variational inference} We have presented a robust pipeline for calibrating potentially misspecified ABMs using generalized variational inference. By leveraging a hybrid AD strategy, we achieve favourable time and space complexity scaling, enabling us to train normalizing flows with many parameters to recover the generalized posterior for ABMs of arbitrary time depth. Furthermore, by using AD tools like \texttt{Bijectors.jl}, we ensure that our optimization-based inference approach produces samples within the ABM parameter domain, maintaining flexibility for any type of ABM parameter.

Our calibration experiments demonstrate that pathwise gradient estimation through differentiable ABMs performs at least as well as score-based methods in variational inference, confirming that the gradients obtained from our differentiable implementations are informative for parameter learning. The performance advantage of pathwise methods becomes more apparent in higher-dimensional parameter spaces, such as the nine-parameter SIR model, compared to simpler models like the Axtell model of firms. This trend aligns with theoretical expectations about the superior scaling properties of pathwise estimators in high-dimensional settings \citep{mohamed_monte_2020}, demonstrating that differentiable ABMs provide a viable approach for gradient-based calibration of ABMs with large parameter spaces.

\paragraph{Julia for differentiable ABMs} All models and experiments were implemented in the \textsc{Julia} programming language \citep{bezanson_julia_2017} \footnote{code available after peer-review}. This choice was motivated by \textsc{Julia}'s multiple dispatch system, which enables seamless switching between AD engines and gradient estimators—a capability that proved essential for this work.

\subsection{Limitations}

\paragraph{Limitations of differentiable ABMs}
The primary limitation of our approach lies in implementation complexity. Constructing differentiable ABMs requires careful consideration of gradient flow through discrete operations and stochastic elements, necessitating thorough testing against finite differences to ensure correct gradient computation. Additionally, the need to maintain gradient information through all possible paths in the computation graph can increase memory requirements and computation time, particularly for models with many discrete choices or large state spaces. Some model parameters, particularly those controlling discrete structural elements like network topology, remain challenging to optimize through gradient-based methods and may require alternative approaches. While we believe the models presented here are widely representative of ABMs, particular models may prove difficult to port to differentiable architectures.

\subsection{Future work}

\paragraph{Computational cost of differentiable ABMs} Despite Julia's computational efficiency, implementation-specific optimizations remain crucial for achieving optimal performance in differentiable ABMs. The models presented in this work have not been optimized for speed and could benefit from several computational advances. In particular, the AD landscape in Julia is rapidly evolving, with packages like \textsc{Enzyme} \citep{moses_instead_2020} offering significant performance improvements over earlier AD systems, which would directly benefit differentiable ABM implementations. Beyond these improvements, \textit{sparse gradient computation} offers promising avenues for computational savings by exploiting the locality principle inherent in most ABMs—computing gradients only for ``sloppy'' parameters that actually influence agents' current states and decisions, potentially reducing computational complexity from $\mathcal{O}(d)$ to $\mathcal{O}(d_{\text{active}})$ where $d_{\text{active}} \ll d$. Additionally, developing memory-efficient reverse mode differentiation for ABMs with extended temporal horizons remains an open challenge.

\paragraph{Advanced gradient estimation techniques}

The gradient estimation landscape for differentiable ABMs requires further theoretical and practical development. The development of adaptive gradient estimators presents a compelling direction, involving methods that automatically switch between Straight-Through, Gumbel-Softmax, and unbiased estimators based on local model characteristics such as the degree of mean-field violation or the sparsity of agent interactions. Such adaptive approaches could dynamically balance bias-variance trade-offs throughout the simulation and could even combine pathwise with score function estimators, leveraging the unbiased nature of score estimators to provide the final precision while using pathwise gradients for the bulk of the optimization trajectory. Furthermore, extending the framework to compute higher-order derivatives would enable second-order optimization methods and more sophisticated uncertainty quantification techniques. Finally, while AD provides efficient local sensitivity analysis, developing methods that bridge local gradient information with global sensitivity methods could provide complementary insights into parameter importance across the entire parameter space.

\subsection{Code Availability}

The implementations of the differentiable ABMs are available at \url{https://github.com/arnauqb/DiffABM.jl} and the calibration interface is available at \url{https://github.com/arnauqb/BlackBIRDS.jl}.


\vskip 0.2in
\bibliography{jmlr}
\section{Connection between ABM and ODE Formulations of SIR Models}
\label{app:sir_ode_connection}

The standard SIR model is formulated as a system of ordinary differential equations,
\begin{align}
\frac{\mathrm{d}S}{\mathrm{d}t} &= -\beta \frac{SI}{N}, \\
\frac{\mathrm{d}I}{\mathrm{d}t} &= \beta \frac{SI}{N} - \gamma I, \\
\frac{\mathrm{d}R}{\mathrm{d}t} &= \gamma I,
\end{align}
where $S(t)$, $I(t)$, and $R(t)$ represent the number of susceptible, infected, and recovered individuals at time $t$, respectively, $N = S + I + R$ is the total population size, $\beta$ is the transmission rate, and $\gamma$ is the recovery rate.

In our ABM formulation, we model state transitions as Poisson processes. A susceptible agent $i$ experiences an infection rate that depends on their local neighborhood,
\begin{equation}
\lambda_i(t) = \beta \sum_{j \in \mathcal{N}(i)} I_j(t),
\end{equation}
where $\mathcal{N}(i)$ represents the set of agents in contact with agent $i$, and $I_j(t) = 1$ if agent $j$ is infected at time $t$, zero otherwise. Since these are Poisson processes, the number of events (infections or recoveries) in a time interval $\Delta t$ follows a Poisson distribution with rate parameter $\text{rate} \times \Delta t$. The probability of no events occurring is $\exp(-\text{rate} \times \Delta t)$, so the probability that at least one event occurs is $1 - \exp(-\text{rate} \times \Delta t)$. This gives us the transition probabilities
\begin{align}
P(S_i \rightarrow I_i) &= 1 - \exp(-\lambda_i(t) \Delta t), \\
P(I_i \rightarrow R_i) &= 1 - \exp(-\gamma \Delta t).
\end{align}

The connection between ABM and ODE formulations becomes precise in the limit of a complete graph, where every agent interacts with every other agent. In this case, the force of infection for any susceptible agent becomes
\begin{equation}
\lambda_i(t) = \beta I(t),
\end{equation}
where $I(t)$ is the total number of infected agents. In the thermodynamic limit ($N \rightarrow \infty$, $\Delta t \rightarrow 0$), the law of large numbers ensures that the stochastic fluctuations in the ABM average out, and the population-level dynamics converge to the deterministic ODE solution,

\begin{align}
\mathbb{E}\left[\frac{\Delta S}{\Delta t}\right] &\rightarrow -\beta \frac{SI}{N}, &
\mathbb{E}\left[\frac{\Delta I}{\Delta t}\right] &\rightarrow \beta \frac{SI}{N} - \gamma I, &
\mathbb{E}\left[\frac{\Delta R}{\Delta t}\right] &\rightarrow \gamma I.
\end{align}

This theoretical connection provides an important additional validation point for our gradient computations. Since the ABM and ODE formulations are equivalent in the complete graph limit, differentiating through the ABM should yield the same gradients as differentiating through the ODE system. We have verified this correspondence by comparing ABM gradients against ODE gradients computed using \texttt{DifferentialEquations.jl} \citep{rackauckas_differentialequationsjlperformant_2017}, providing additional confidence in our differentiable ABM framework beyond the finite differences comparison presented in \autoref{sec:sir_gradients}.
\section{Additional experimental details}
\label{app:exp_details}

\subsection{Fixed Parameters}

\paragraph{Axtell Model of Firms} We employ $N=2000$ agents and generate a contact graph by randomly sampling 1 to 4 neighbors for each agent. The model runs for 30 time-steps.

\paragraph{Sugarscape} We simulate $N=100$ agents on a $25\times 25$ toroidal grid for 50 time-steps. The sugar regeneration rate is fixed at $r=1$. Agent wealth is sampled as $6 + 19 \times \text{Beta}(w_\alpha, w_\beta)$ to ensure initial wealth ranges between 6 and 25. Metabolism is sampled as $2 + 2 \times \mathrm{Beta}(m_\alpha, m_\beta)$.

\paragraph{SIR model} We use $N=2000$ agents with an Erdős–Rényi graph having edge probability $p=10^{-2}$, resulting in approximately 20 neighbours per agent on average. The model runs for 60 time-steps with $\Delta t=1$ hour.

\subsubsection{Models free parameter values}

\autoref{tab:true_params} shows the values chosen for the free parameters of each model across our experiments, including both the calibration and comparison to FD experiments. 

\begin{table}[htbp]
\centering
\begin{tabular}{ll|ll}
\toprule
\multicolumn{2}{c|}{\textbf{SIR Model}} & \multicolumn{2}{c}{\textbf{Sugarscape Model}} \\
\textbf{Parameter} & \textbf{Value} & \textbf{Parameter} & \textbf{Value} \\
\midrule
$\log_{10}(I_0)$ & $\log_{10}(10^{-2})$ & Vision prob. ($p_1$, $p_6$) & [0.2, 0.8] \\
$\log_{10}(\beta)$ & $\log_{10}(0.4)$ & Metabolic ($m_\alpha$, $m_\beta$) & $[2.0, 5.0]$ \\
$\log_{10}(\gamma)$ & $\log_{10}(0.05)$ & Wealth ($w_\alpha$, $w_\beta$) & $[5.0, 2.0]$ \\
Soc. dist. start ($D_\mathrm{start}$) & $10.0$ & & \\
Soc. dist. end ($D_\mathrm{end}$) & $45.0$ & \multicolumn{2}{c}{\textbf{Axtell Model of Firms}} \\
Soc. dist. strength ($\alpha_D$) & $0.3$ & \textbf{Parameter} & \textbf{Value} \\
Quarantine start ($Q_\mathrm{start}$) & $20.0$ & Income-leisure ($\theta_{\alpha}$, $\theta_{\beta}$) & $[1.0, 3.0]$ \\
Quarantine end ($Q_\mathrm{end}$) & $35.0$ & Initial effort ($e_{\alpha}$, $e_{\beta}$) & $[2.0, 1.0]$ \\
Quarantine prob. ($p_Q$) & $0.7$ & Production scale ($a_{\alpha}$, $a_{\beta}$) & $[2.0, 5.0]$ \\
& & Return increase ($b_{\alpha}, b_{\beta}$) & $[5.0, 2.0]$ \\
\bottomrule
\end{tabular}
\caption{True parameter values for synthetic data generation}
\label{tab:true_params}
\end{table}

\subsection{Calibration}

\subsubsection{Normalizing Flow's architecture}

We use a masked affine autoregressive normalizing flow \citep{huang_neural_2018} from the \textsc{normflows} Python's library \citep{stimper_normflows_2023}. We have built a Julia wrapper around the Python library to allow automatic differentiation using \textsc{Zygote.jl} \citep{innes_dont_2018} \footnote{Code available at [anonymous]}.

The autoregressive flow contains 4 layers, each with 2 blocks of 32 hidden units and a linear permutation across dimensions. The base distribution is a multivariate normal. 

\subsubsection{Training hyperparameters}

We use the AdamW optimizer \citep{loshchilov_decoupled_2019} with running average coefficients $\beta_1=0.9$ and $\beta_2=0.99$. The gradient is normalized to unit length on each step (norm clipping). We employ 5 Monte-Carlo samples per step. The flow trains for 5,000 epochs without early stopping and the weights with the lowest validation loss across all epochs are saved.

We use the maximum mean discrepancy loss (MMD) \citep{gretton_kernel_2012} to evaluate model fit with a sample size of 2.
\end{document}